\RequirePackage{fix-cm}

\documentclass[smallextended]{svjour3}       
\smartqed  
\usepackage{graphicx}
\usepackage{booktabs} 

\usepackage[ruled]{algorithm2e} 

\usepackage[utf8]{inputenc}
\usepackage{amsmath,amssymb,natbib,mathtools,booktabs,floatrow,caption, subcaption, bbm, bm, chngcntr}
\usepackage[normalem]{ulem}
\usepackage[ruled]{algorithm2e}
\usepackage[section]{placeins}
\usepackage[english]{babel}
\usepackage{authblk}
\usepackage{graphics}
\usepackage{graphicx}
\usepackage{mathptmx}
\usepackage[toc,page]{appendix}
\usepackage{geometry}
\usepackage{lipsum}

\newcommand{\ignore}[1]{}

\usepackage{natbib}
 \bibpunct[, ]{(}{)}{,}{a}{}{,}%
 \def\bibhang{24pt}%

\begin{document}

\title{Displaying Things in Common to Encourage Friendship Formation: \\A Large Randomized Field Experiment
}

\titlerunning{Displaying Things in Common to Encourage Friendship Formation}        

\author{Tianshu Sun         \and
        Sean J. Taylor 
}


\institute{Tianshu Sun \at
              Marshall School of Business, University of Southern California\\
              \email{tianshus@marshall.usc.edu}           
           \and
           Sean Taylor \at
           Facebook \\
                \email{seanjtaylor@gmail.com}
}

\date{}

\maketitle

\begin{abstract}
Friendship formation is of central importance to online social network sites and to society, but can suffer from significant and unequal frictions. In this study, we demonstrate that social networks and policy makers may use an IT-facilitated intervention -- displaying things in common (TIC) between users (mutual hometown, interest, education, work, city) -- to encourage friendship formation, especially among people who are different from each other. Displaying TIC may update an individual's belief about the shared similarity with another and reduce information friction that may be hard to overcome in offline communication. In collaboration with an online social network, we design and implement a randomized field experiment, which randomly varies the prominence of different types of things in common information when a user (viewer) is browsing a non-friend's profile. The dyad-level exogenous variation, orthogonal to any (un)observed structural factors in viewer-profile's network, allows us to cleanly isolate the role of individuals' preference for TIC in driving network formation and homophily. We find that displaying TIC to viewers may significantly increase their probability of sending a friend request and forming a friendship, and is especially effective for pairs of people who have little in common. Such findings suggest that information intervention is a very effective and zero-cost approach to encourage the formation of weak ties, and also provide the first experimental evidence on the crucial role of individuals' preference (versus structural embeddedness) in network formation. We further demonstrate that displaying TIC could improve friendship formation for a wide range of viewers with different demographics and friendship status, and is more effective when the TIC information is more surprising to the viewer. Our study offers actionable insights to social networks and policy makers on the design of information intervention to encourage friendship formation and improve the diversity of the friendship, at both an aggregate and an individual level.
\end{abstract}

\clearpage

\section*{1. Introduction and Motivation}
\label{intro}
Friendship formation is of central importance to online social network sites (SN) and to society. For social networks such as Facebook, Instagram, and Linkedin, encouraging friendship formation among users is vitally important: first, articulated friendships provide infrastructure for the spread of information \citep{granovetter1977strength,yoganarasimhan2012impact,peng2018network}, influence \citep{aral2011creating,bapna2015your,ugander2012structural,eckles2016estimating}, and support \citep{burke2016relationship} on the network. In addition, user engagement and retention on SN sites exhibits a strong local network effect \citep{sundararajan2007local,hartmann2010demand,tucker2008identifying} and friendship formation is primarily what enables this effect: with more friends, a user may use the SN more frequently, share more and create more content \citep{shriver2013social,goes2014popularity,toubia2013intrinsic}. Recognizing the significant benefits users gain from having more friends, most major SN sites invest significantly in encouraging friendship formation among their users\footnote{Several major SNs have made increasing the number of friends a key goal in their operations \citep{price2012growthhacking}: e.g. Linkedin aims to get a user to reach X friends in Y days, and Facebook and Twitter use similar goal metrics.}. 
The same considerations and emphasis on friendship formation have also been prevalent among public and private organizations in the society: policy makers in schools and corporations recognize the importance of friendship for the stability and performance of the students and employees \citep{marmaros2006friendships,carrell2013natural,mayer2008old,mollica2003racial,ibarra1992homophily}, and strive to facilitate tie formation within their organizations. 

Besides the absolute volume of friendship formation among individuals, there is also a surging interest and an increasing emphasis by both SN sites and policy makers on the diversity of friendships \citep{fischer2008does,eagle2010network}. A diverse friendship, formed by weak ties (i.e. people who share little in common in terms of mutual friends, origin, past experience, as defined in \cite{easley2010networks,aral2014tie}) may increase information novelty \citep{bakshy2012role}, facilitate the spread of valuable information \citep{gee2017paradox,shi2013content}, and promote mutual understanding among different parties \citep{bakshy2015exposure}. 

Despite the importance of the volume and diversity of friendships, there are significant frictions in the friendship formation process (e.g. information friction) and these frictions are unequally distributed -- they are often higher among people who are different from each other. Previous research in Marketing, Information Systems, Economics and Sociology have been focused on understanding the preference and mechanisms underlying network formation \citep{katona2008network,jackson2008social,mayzlin2012link,ameri2017structural,phan2015natural,mcpherson2001birds,altenburger2018monophily}, but no systematic research has been done on how policy makers can actively encourage friendship formation, especially for people who are very different (with little in common). In this study, we propose to address the challenge using a novel IT-facilitated information intervention: displaying `things in common' computed from big data of users’ digital profile on social networks (origin, past experience, likes, check-ins). 

The IT-facilitated TIC intervention is motivated by theories on friendship formation and enabled by recent advances in big data and computing technology. When making a friending decision, individuals search for shared attributes or experiences with the other party \citep{fisman2008racial,currarini2009economic}, such as common origin (hometown), experiences (work, education), contexts (current city), or interests (likes). Information about TIC can help reduce cognitive cost \citep{mollica2003racial,forman2008identity}, increase trust \citep{lin2015home}, and allow individuals to derive more utility from the friendship in the long run \citep{felmlee1990dissolution}. However, discovering similarity is often costly and suffer from informational friction in offline communication: it may take multiple conversations and a non-trivial amount of time before two people realize they share something in common. In contrast, with recent development in data infrastructure and technology, online social networks can now compute, synthesize and display things in common between any pair of users in real time (when a viewer is browsing a non-friend profile). Displaying TIC updates the viewer’s belief about shared similarities with another\footnote{The viewer may either have little information about the profile's interest, or have an biased perception\cite{goel2010real}. In both scenarios, the displayed TIC information would help update the viewer's belief.} and can significantly reduce information friction in the friending process, generating substantial value in network formation. 

Interestingly, such information intervention we suggest can be applicable in any IT-mediated friendship formation process in both online social platforms and offline organizations (e.g. through enterprise social network like Slack, Workplace, Jive and Yammer). And the potential of TIC intervention is only increasing
because of the accumulation of rich data and the advances of computing technology  \citep{yadav2014marketing}. To the best of our knowledge, our study is the first to identify the unique opportunity and propose the information intervention in both industry practice and academic research. The promise of displaying TIC is huge: as envisioned by Zuckerberg in Facebook's new mission statement ``If we connect with people about what we have in common, it is easier to have dialogue about what we disagree on. When we do this well, we give billions of people the ability to share new perspectives...and bring the world closer together'' \citep{zuckerberg2017bringing}

Despite the practical importance and technical feasibility, no study has systematically examined the effectiveness of displaying things in common on friendship formation. Very little is known about its causal effect on users' friending behaviors, friendship formation, and friendship diversity. Specifically, we want to understand:

(Q1, effectiveness) Can displaying `things in common' encourage a viewer's friend request and increase the probability of friendship formation? 

(Q2, heterogeneity) Is displaying TIC effective in encouraging the friendship formation of weak ties, i.e. pairs of individuals who are different (with no mutual friends and who share few TICs)? Is displaying TIC effective for viewers with different demographics and friendship status?

(Q3, mechanism) What is the potential mechanism underlying the effect of displaying TIC on viewers' friending decision? When will the display of a TIC be more effective? Will there be substitution or complementarity when showing multiple TICs and why?

The answers to these questions provide immediate \textbf{managerial} implications on how to facilitate users’ friending decision and cultivate tie formation in social networks -- a policy question that is important not just for large SN and organizations but for any site or firm building social features. 

To address the questions, we carefully design and conduct a randomized experiment at the level viewer-profile pairs, and find that displaying TIC may significantly increase friendship formation, especially for weak ties (by 7.1-10.3\%). We demonstrate that TIC intervention works for a wide range of users and is more effective when TIC creates more `surprisal' to the viewers. As a result, the proposed TIC intervention is implemented at full scale. In this way, our experiment contributes to the practice and the Marketing, Information Systems and Economics literature, in few ways.

First, a large stream of literature on social influence (or peer effect) have intensively studied IT-mediated social influence on existing friendship ties \citep{iyengar2011opinion,tucker2008identifying,aral2012identifying,susarla2012social,wang2018socially}. However, little is known about how IT can help facilitate and shape friendship formation. Understanding IT-enabled tie formation is crucial as friendship ties serve as the infrastructure underlying influence processes and have a significant impact on information diffusion \citep{shi2013content,peng2018network}. Our study is among the first to address this gap and suggests that SN sites may creatively leverage IT-facilitated information intervention to overcome challenges in friendship formation, especially among weak ties\footnote{Despite lively discussions on the importance of friendship diversity \citep{kossinets2009origins,eagle2010network,granovetter1977strength} in facilitating interactions among different groups \citep{bakshy2015exposure}, very little is understood about how online social networks can actively help `build' the weak ties among people with different background.}. In this way, information and influence can better spread on the network infrastructure \citep{phan2018evolution,gee2017paradox}. 

Second, recent efforts by researchers, policy makers and SN sites to improve friendship formation have focused on `recommending the right people' for interaction (e.g. `People You May Know' feature, \citep{linkedin2016pymk,su2016effect,brzozowski2011should}. 
In this study, we argue that it is equally important to `show the right information' upon interaction. Regardless of the friend recommendation algorithm, a user will spend a limited amount of time browsing another user’s profile on Facebook and Twitter (\cite{lerman2013limited}, few seconds). Providing better information to viewers allows them to understand what they share in common with other users and make more informed friending decisions. In addition, unlike friend recommendation/link prediction algorithms which generally rely on the existence of mutual ties \citep{moricz2010pymk,dhar2014prediction}, information provision can operate independently of social network structure thus can help connect people who share no mutual ties (i.e. weak ties formation).\footnote{Finally, our study also extends a large stream of literature in Marketing and IS on the reduction of information friction in online platforms and marketplaces. Previous literature has focused on how IT artifacts (online reviews, product recommendations) can be used to reduce friction in user-product interactions \citep{fradkin2017search,forman2008identity,oestreicher2012recommendation,fleder2009blockbuster}. Our study suggests a new route in which IT could reduce the friction in user-user interaction, and open up a new area of research on the role of IT in moderating the structure, evolution and value of user-user network \citep{oestreicher2013network,hosanagar2013will}.}  

Third, despite a recent surge of interest in diversity at schools and workplaces and an increasing emphasis on friendship diversity \citep{fischer2008does}, little is understood about how platforms and policy makers can foster such diversity on platforms or within organizations. Our study proposes and demonstrates that displaying TIC information, distilled from existing data accumulated on the platform (or in the organization), could reduce information friction in social interaction process and increase friendship diversity. We envision that TIC information may also reduce bias in other forms of social interaction such as hiring and increase diversity within an organization.

Besides managerial relevance and immediate applicability, our study also makes several contributions to social network literature and improve our \textbf{theoretical} understanding of the origin of network formation and homophily. First, our experiment cleanly isolates the causal effect of users' preferences for different types of TIC on tie formation. This is the first experimental evidence in the network literature. Second, the origin of homophily is notoriously hard to identify \citep{currarini2010identifying}, as the preference factor (over TIC) and structural factor (e.g. network embeddedness or mutual friends) always entangles with each other in observational data. By randomizing the display of TIC (which is orthogonal to any structural factor), we are now able to measure the relative importance of individuals' preferences versus structural factors in network formation process. We further detail the contributions to network theory in `Results and Discussion' and in Appendix D.

\section*{2. Identification Challenges and Strategy}
\label{sec:identification}

\begin{figure}[h!]
\centering
\includegraphics[width=0.65\textwidth]{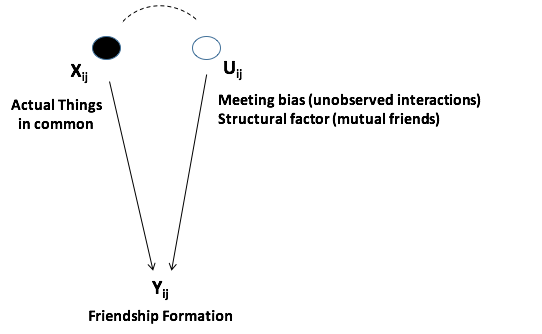}
\caption{\emph{Identification Challenge.} In observational data, it is extremely hard to identify the effect of displaying TIC in driving friendship formation (${Y}_{ij}$), as the actual TIC (${X}_{ij}$) for viewer-profile pair is endogenously correlated (dashed line) with confounding factors (${U}_{ij}$) such as the existence of mutual friend, thus the estimated effect of ${X}_{ij}$ on ${Y}_{ij}$ is biased. (formally, in econometric equation ${Y}_{ij}= \gamma*{X}_{ij} + {U}_{ij}, \gamma $ is biased)}
\label{fig:iden_challenge}
\end{figure}

\begin{figure}[h!]
\centering
\includegraphics[width=0.75\textwidth]{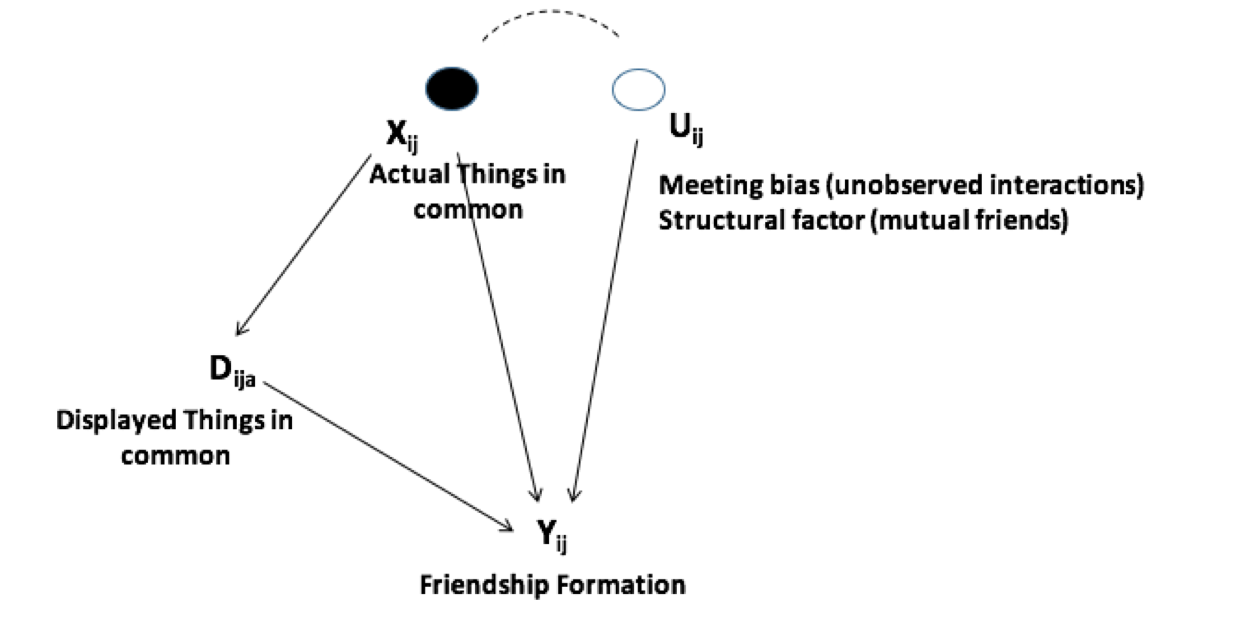}
\caption{\emph{Identification Strategy.} Randomization process in the experiment creates exogenous variation in the displayed things in common (${D}_{ij}$), therefore allows us to identify the causal effect of showing TIC and isolate the effect of displaying TIC in driving friendship formation. The display of a certain type of things in common `a' (i.e. ${D}_{ija}$) is independent of ${U}_{ij}$, thus the estimated coefficient for ${D}_{ija}$ is unbiased. The exogenous assignment of ${D}_{ija}$ allows us to identify effect of each type of things in common. (formally, ${Y}_{ij} =  {\sigma}_{a}*{D}_{ija}  + \gamma*{X}_{ij}  + {U}_{ij}$, where ${\sigma}_{a}$ is the causal effect of displaying a type of TIC `a'. ${\sigma}_{a}$ is identified and unbiased because ${corr({D}_{ija}, {U}_{ij})=0}$).}
\label{fig:iden_experiment}
\end{figure}

While important to both practice and theory, identifying the causal effect of displaying things in common on friendship formation is empirically challenging (Figures~\ref{fig:iden_challenge}~and~\ref{fig:iden_experiment}). First, one cannot identify the treatment effect by simply comparing friending decisions across ego-alter pairs with different numbers of TIC: it is possible that the ego may be more likely to friend an alter that shares more similar traits due to unobserved confounding (${U}_{ij}$ in Figure~\ref{fig:iden_challenge}, e.g. existence of mutual friend), regardless of whether the `things in common' are actually shown on the alter profile or not. In other words, when observing a correlation between the presence of TIC (i.e. ${X}_{ij}$) and an increase in friendship formation (i.e. ${Y}_{ij}$), one can not differentiate structural factors versus treatment effect (of displaying TIC)\footnote{For instance, those viewer-profile pairs with 2 TIC may have a higher friendship formation rate than those with only 1 TIC, not because of the additional TIC at display, but because pairs with 2 TIC are likely to have more mutual friends and more interaction opportunities (i.e. structural factors) than pairs with 1 TIC. In Online Appendix C, we confirm the above insight and empirically demonstrate that the correlation between number of things in common in a pair and the corresponding friendship formation rate is not only biased, but even opposite to the true causal effect in sign (Figure~\ref{fig:corr_vs_causal})}. This identification problem is fundamentally rooted in unobserved confounding present in observational data \citep{hartmann2008modeling,phan2015natural,kossinets2009origins,aral2009distinguishing, shalizi2011homophily,kwon2017platform}\footnote{Observational data often lacks of detailed information on the structural factor such as interaction history and friendship structure \citep{mcpherson2001birds}, which is needed to control for meeting bias and triadic closure. Even more fundamentally, the endogenous correlation between things in common, meeting bias, and network structure makes it almost impossible to isolate the role of preference in friendship formation from observational studies \citep{currarini2010identifying}.}. 
Second, 
structural factor may complement or substitute the effect of preferences over TIC in the network formation process. It would be interesting to understand the importance of displaying TIC with or without the presence of mutual friends. 
Finally, the baseline probability of friendship formation in online social interaction process is generally very low. Less than 0.1\% of the viewer-profile pairs form a friendship in the social interactions (viewer browsing the profile). Thus, a very large sample of pairs is needed to detect statistical significance of a reasonable treatment effect \citep{lewis2015unfavorable}.

In order to address these challenges, we leverage a large-scale randomized experiment on a large SN site involving over 50 million viewer-profile pairs.  This product test was designed to help choose the best combination of TIC for the SN site to display, randomly varying the prominence of ‘things in common’ information when a viewer is browsing another’s profile card. The treatment (to display certain TIC information or not) is randomly assigned only thus are independent to both observed and unobserved structural factor between the viewer-profile dyad (e.g. mutual friend(s), interaction opportunity). The exogenous variation created in the displayed TIC (${D}_{ij}$) boosted the salience of TIC to the viewer and allows us to identify the role of preference in driving friendship formation (Figure~\ref{fig:iden_experiment}), and understand its effect with or without the presence of structural factors (i.e. the existence of mutual friend) in a real network environment. The scale of our experiment ensures that we have sufficient statistical power to detect the treatment effect, and to explore heterogeneity in effects for different categories of TIC and for different subgroups of viewer-profile pairs. We present the experiment and data in detail as below.

\section*{3. Experiment Design and Data}
We collaborated with a large global SN site which features a user profile page where registered users can report a wide range of information that they want to share with others, including current city, current work, hometown, past education, past experience and so on. Users on the SN site can also participate in several kinds of activities, including browsing and interacting with friends’ updates and public pages, and posting updates and content on their own profile. 

\begin{figure}[h!]
\centering
\includegraphics[width=0.7\textwidth]{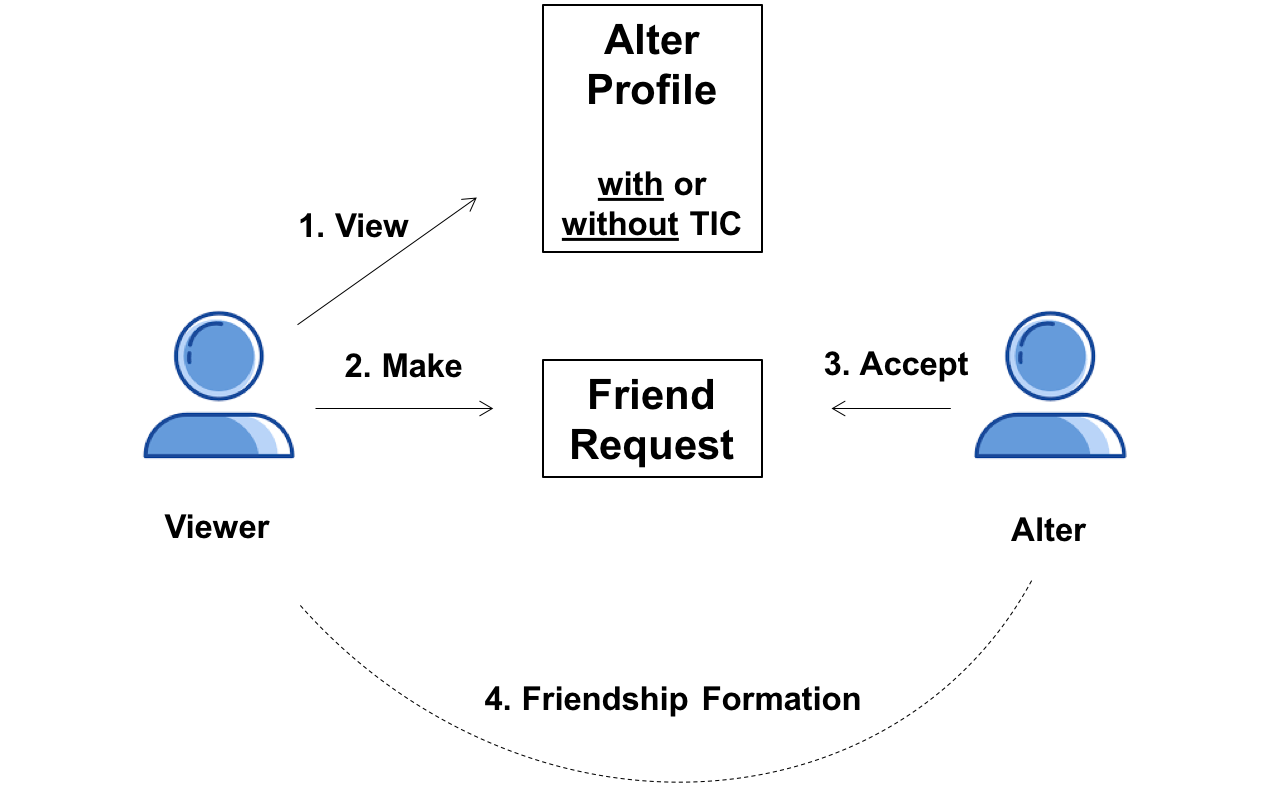}
\caption{\emph{Process.} Step-by-step illustration of friendship formation process.}
\label{fig:process}
\end{figure}

\begin{figure}[h!]
\centering
  \begin{subfigure}[b]{0.40\textwidth}
    \includegraphics[width=\textwidth]{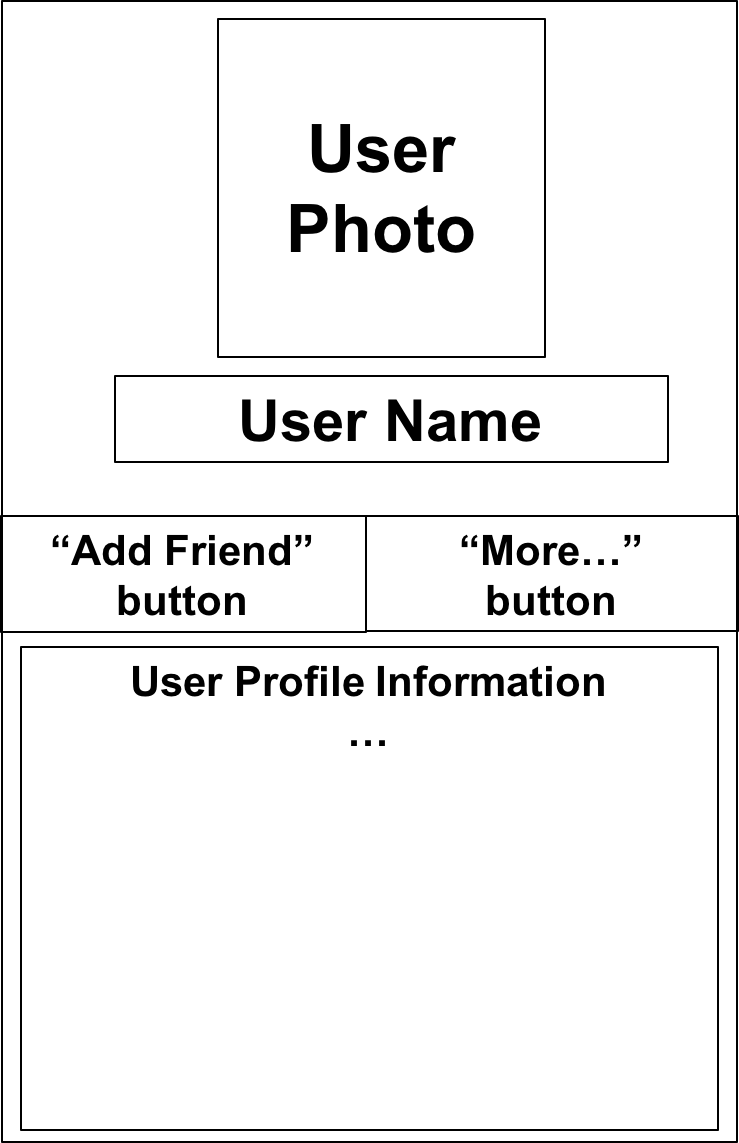}
    \label{fig:wire_control}
  \end{subfigure}
  \begin{subfigure}[b]{0.40\textwidth}
    \includegraphics[width=\textwidth]{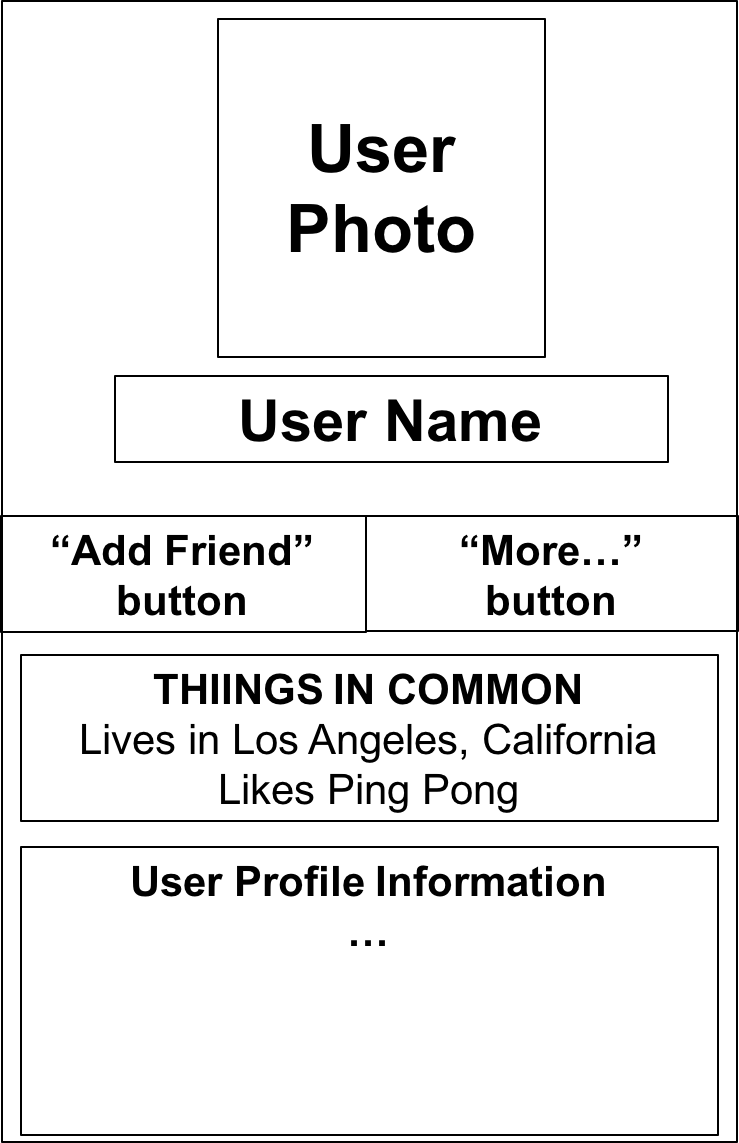}
    \label{fig:wire_treatment}
  \end{subfigure}
\caption{\emph{Control and Treatment Group.}  A wireframe of the user interface for the experiment. Control group is on the left and treatment group on the right. In this example, the two types of TIC between the profile and the viewer -- mutual city and mutual like  -- are both displayed}
\label{fig:wire}
\end{figure}

The SN site, like most other SN sites, has a function on its mobile platform in which users can discover new friends. Specifically, a user (who we will refer to as the 'viewer') may view the profile cards of other users -- one at a time on the mobile screen (Figure~\ref{fig:wire}). The profile card includes a subset of information that the alter has shared on her or his public profile on the social network site, including current work, education, current city, hometown, as well as recent photos and updates s/he has shared publicly (Figure~\ref{fig:wire} for a wireframe of the design). The viewer may swipe right or left to see the next or previous profile card in the sequence, and can tap on “Add Friend” to send a friend request to the alter. If the alter approves the request in the notification, then an articulated friendship between the pair is formed (Figure~\ref{fig:process})\footnote{We use 'articulated friendship' to denote that the friendship on SN sites is a unique type of social network connection, which is valuable by itself (in the creation and spread of information) and may differ from the offline friendship. For instance, the interaction frequency and tie strength of articulated friendship on average may be lower (weaker) as compared to the offline friendship.}. When browsing the profile card for a specific alter, the viewer may also see whether s/he share any mutual friends with the profile, and if so, how many. Finally, the site may also display things in common for the viewer-profile pair (Figure \ref{fig:wire} right pane) based on their public profile information (city, work, hometown, education) and page like behavior in the past\footnote{Specifically, the viewer and profile can be represented as two high-dimensional $n\times1$ vectors: each row representing the value of a profile field entry or a page like decision. Their TIC are calculated from the intersect of the two vectors.}. There are five types of things in common: mutual city, mutual work, mutual hometown, mutual education, and mutual interest (page likes). Depending on the characteristics and past behaviors of both the viewer and profile, each viewer-profile pair may have zero, one, or up to five things in common. Thus, the eligible things in common that the SN site can display for each viewer-profile pair may vary across pairs. 

\begin{figure}[h!]
\centering
\includegraphics[width=0.7\textwidth]{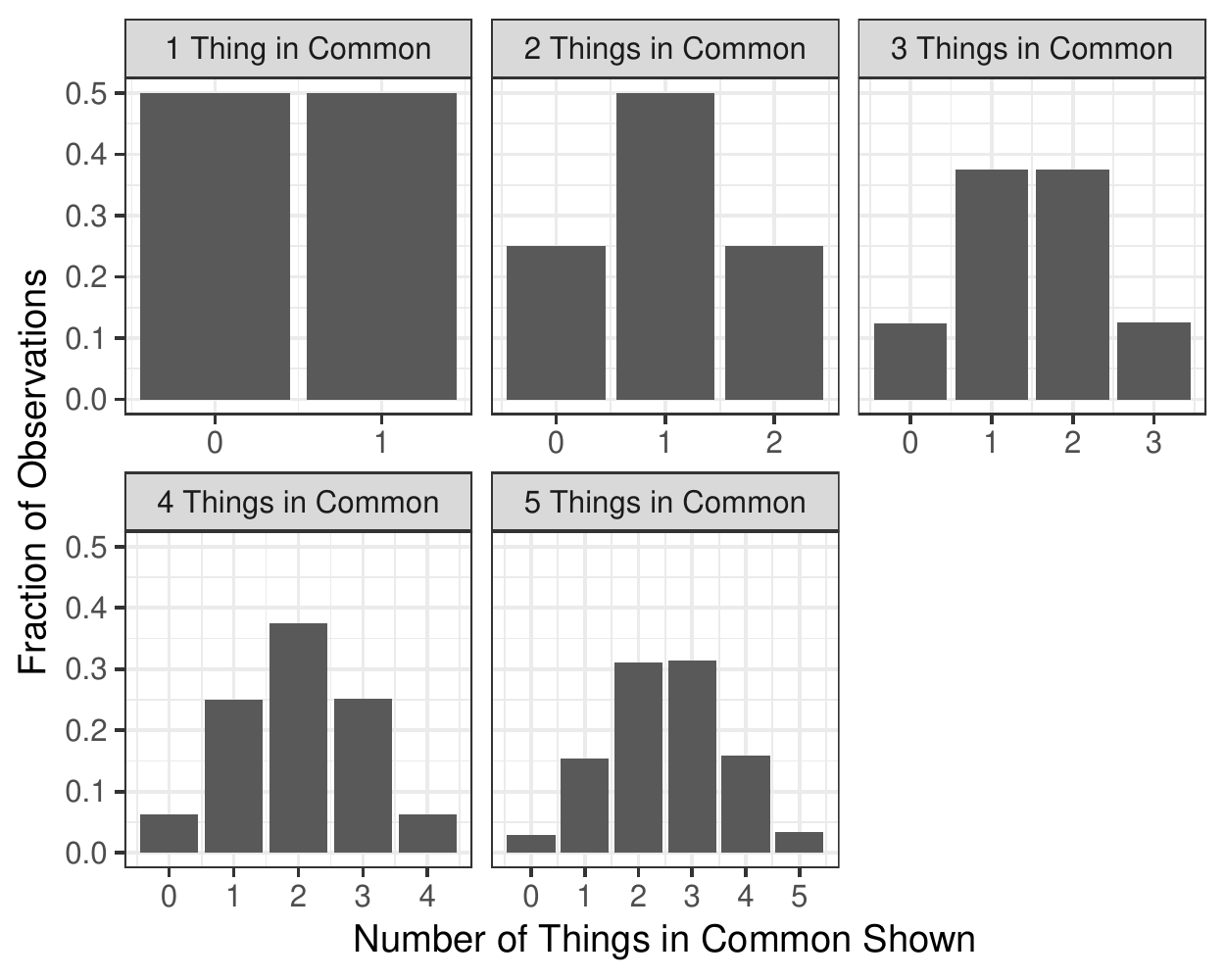}
\caption{\emph{Randomization.} Number of TIC shown to the viewer follows a binomial distribution, grouped by number of actual TIC shared between the pair. The empirical distribution perfectly matches our randomization strategy, providing the randomization check.}
\label{fig:randomization}
\end{figure}

We randomly sampled a portion of viewer-profile pairs with at least one TIC (and at most five TIC) and conduct a randomized field experiment over the representative sample. Over a period of few weeks, over 50 million viewer-profile pairs were randomly assigned into different test conditions. Specifically, for each viewer-profile pair, we randomly and independently vary whether to display each type of things in common that the pair may share, with a 50\% chance of showing or not (Step 1 in Figure~\ref{fig:process}). For instance, for all pairs with two TIC (e.g. mutual city and mutual like), in expectation 25\% of pairs would see two TIC, 50\% would see one (with 25\% see mutual like and 25\% see mutual city), and 25\% would see none. In expectation, the number of TIC shown should follow a binomial distribution determined by number of actual things in common shared by the viewer-profile pairs. Figure~\ref{fig:randomization} confirms the empirical distribution perfectly matches our randomization strategy, providing the randomization check\footnote{We also check the balance of a series of user covariates (e.g. viewer/profile gender, friend count) across the control and treatment groups and do not find any significant differences. All tables are available upon request.}. Finally, the randomization is performed at viewer-profile pair level, i.e. the same viewer may be assigned to different condition when browsing different profiles.  We choose to randomize at viewer-profile level because the primary outcomes of interest, friend request and friendship formation, are measured at that level.

Besides the randomized display of TIC, we also controlled the information environment of viewer-profile interaction in our experiment in two ways: first, upon the viewer sending out a friend request, the alter in both control and treatment group will see exactly the same notification (Step 3 in Figure~\ref{fig:process}), and no information about TIC is displayed to the alter in the notification. In this way, we isolate the treatment to viewers and all effects of displaying TIC on friendship formation should be driven by changes in viewers' decision. Second, the profile sequence that the viewers browse was generated in the same way across the control and treatment groups, and the test group is only assigned after the creation of profile sequence for each viewer. Such additional efforts help us isolate the effect of displayed TIC on the friendship formation process and eliminate confounding.

The randomized experiment creates exogenous variation so that the displayed TIC in a pair a) is decoupled from actual TIC (i.e. exogenous variations of displayed TIC within each group in Figure 5), and b) is independent of the interaction history and network structure for any viewer-profile pair. As a result, any unobserved confounding factors that may drive friendship formation for a viewer-profile pair (Figure~\ref{fig:iden_challenge}), such as unobserved interaction \citep{currarini2009economic} and structural factor (sharing mutual friend or not), are orthogonal to treatment assignment. Any difference in the friendship formation between viewer-profile pairs in control and treatment groups (with the same actual things in common) can be attributed to difference in the displayed things in common. The exogenous variation created in the experiment allows us to cleanly identify the causal impact of showing things in common on friendship formation, therefore also isolate the role of preference over shared TIC in driving network formation.

For every viewer-profile pair in our experiment, we log 1) the actual things in common between the pair, 2) things in common shown on the profile card (i.e. test group assignment), 3) whether they share mutual friends and 4) some limited demographic information for both viewer and profile (gender, city, etc.), as well as 5) the outcomes of the interaction (e.g. friend requests made by the viewer and friend requests accepted by the profile’s associated user). With the experiment data, we can cleanly identify the causal effect of displaying ‘things in common’ on friendship formation, as well as its impact across different types of things in common, on different types of viewers, and for weak ties versus strong ties. 

We first present the summary statistics for our experiment in Table \ref{tab:summary_stats} and Figures. As shown in Figure \ref{fig:ntics_dist}, the vast majority ($>$98\%) of the viewer-profile pairs in our sample only have one or two things in common (similar pattern when we separately draw the distribution for pairs with and without mutual friend, Figure \ref{fig:ntics_dist_grouped}). Thus we will focus on discussing the treatment effect for those pairs. Among all the pairs with one TIC, the viewers and profiles are most likely to share mutual likes, followed by mutual hometown, city, education and work (Figure \ref{fig:ntics_1tic}). Pairs with two TIC follows a similar ranking (Figure \ref{fig:2tics_dist}). In total, only 53.3\% of viewer-profile pairs in our experiment share at least one mutual friend (Table \ref{tab:summary_stats}) thus 46.7\% of pairs are weak ties. 

On average, a viewer may browse multiple profiles during the experiment period (Figure~\ref{fig:dist_profile_viewer} left pane) but a profile is very likely be browsed by only one viewer (Figure~\ref{fig:dist_profile_viewer} right pane). Since our randomization at viewer-profile level, we sometimes sample the same viewer multiple times as s/he may browse multiple profile cards. To account for dependence introduced by repeated observations from the same viewer in our sample, we use a clustered bootstrap, randomly re-sampling 50\% of viewers without replacement 500 times and deriving 95\% confidence intervals from the sampling variance of the bootstrap replicate means \citep{owen2012bootstrapping}. Bootstrapping for inference is a common practice in the analysis of experiments with dependence between observations. \cite{bakshy2013uncertainty} provide detailed simulations of how dependence in experimental data affect statistical inference in experiments, showing that bootstrapping is more conservative than a variance estimator (that assumes independence) and reduces the probability of Type I errors\footnote{Treatment effect from variance estimators yields the same point estimate and more statistically significant results.}. 

\section*{4. Results and Discussions}
We present the findings from the experiment in this section. We organize the results and discussions based on the managerial and theoretical perspective. The managerial perspective corresponds to our research question Q1 and Q2, while the theoretical perspective corresponds to Q3.  
\subsection*{\textbf{4.1. Managerial}}
\begin{figure}[h!]
\centering
\includegraphics[width=0.8\textwidth]{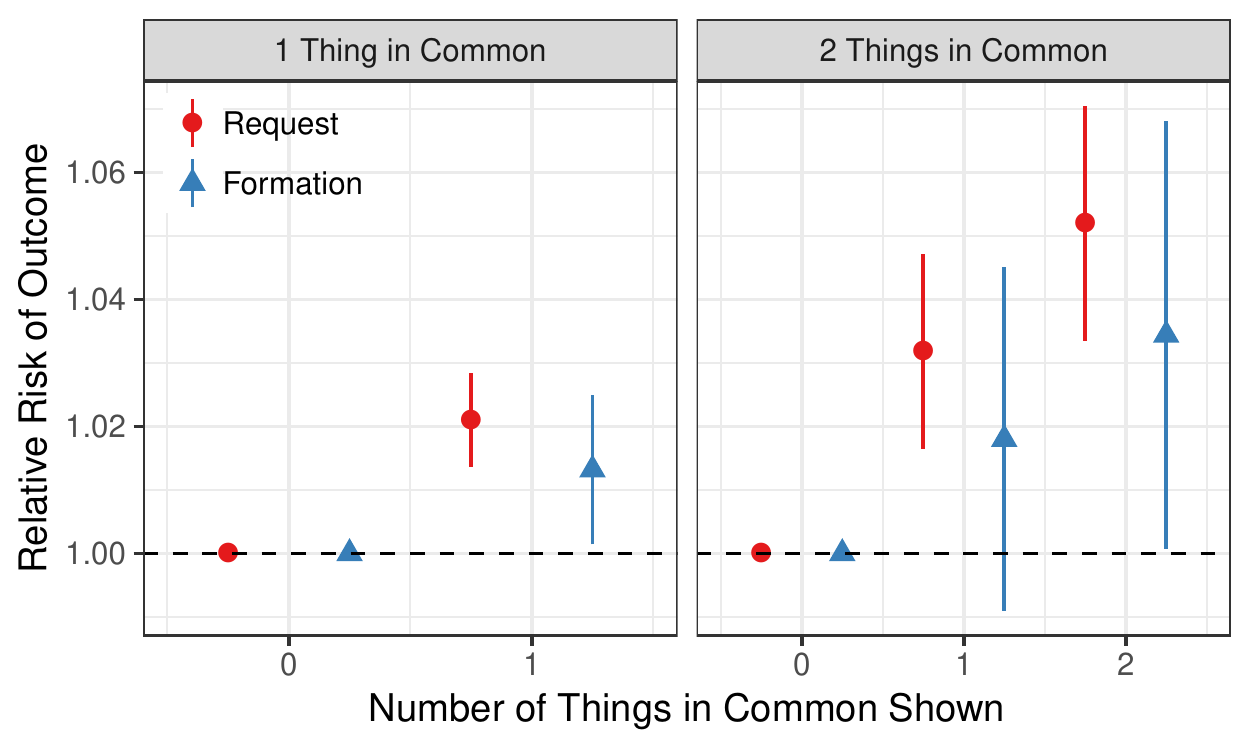}
\caption{\emph{Effects of displaying things in common.} Using pairs with zero things in common displayed as a baseline, showing more things in common increases the relative risk of both friendship request and formation. The effect are statistically significant for pairs with 1 and 2 actual things in common (left and middle pane). 95\% confidence intervals are estimated using a bootstrap clustered on viewer with $500$ bootstrap replicates.}
\label{fig:ntics}
\end{figure}

Our experiment generates three main findings, which may provide immediate managerial implications to social network sites and policy makers. First (Q1), we find that showing things in common significantly increases friendship formation. As shown in Figure \ref{fig:ntics}, within each group with the same number of actual TIC (1, 2 or 3), displaying TIC to the viewers would lead to a significant lift in their likelihood of sending out a friend request. The relative lift, as compared to the control group within each TIC group, ranges from 2.1\% for pairs with 1 TIC to 8.9\% for pairs with 3 TIC. It suggests that providing information about things in common may facilitate users' decision in friending process. Consistent with the pattern on friend request, we find displaying all things in common may significantly increase the probability of friendship formation\footnote{Same pattern holds for pairs with 3 actual TIC (Figure \ref{fig:ntics_3}, \ref{fig:ntic_grouped_3}) but sample size is too small as shown in Figure \ref{fig:ntics_dist}.}: a 1.3\% increase in relative risk for pairs with 1 TIC and a 3.5\% increase in relative risk for pairs with 2 TICs (we present the detailed estimates on the effect for both request and formation and the statistical significance in Table \ref{tab:effects1}). The increase in friendship formation is largely driven by additional friend request by the viewer (Step 2 in Figure \ref{fig:process}), as revealed from the similar lift in the request and formation outcome in Figure \ref{fig:ntics}\footnote{We are unable to find any evidence that the treatment induces any significant positive or negative effects on the rate that requests are accepted (see Figure \ref{fig:conditional_acceptance_rate} for detailed analysis).}. Such an increase is significant given the very large volume of viewer-profile pairs in the browsing session every day. Finally, we also compare the `quality' of friendship ties formed in control and treatment groups, as measured by follow-up interactions (e.g. like, comment) within the ties. We find the IT-enabled friendship ties (with TIC display) is as good as regular ties in facilitating future interactions (Figure \ref{fig:post_interact}).

\begin{figure}[h!]
\centering
\includegraphics[width=0.8\textwidth]{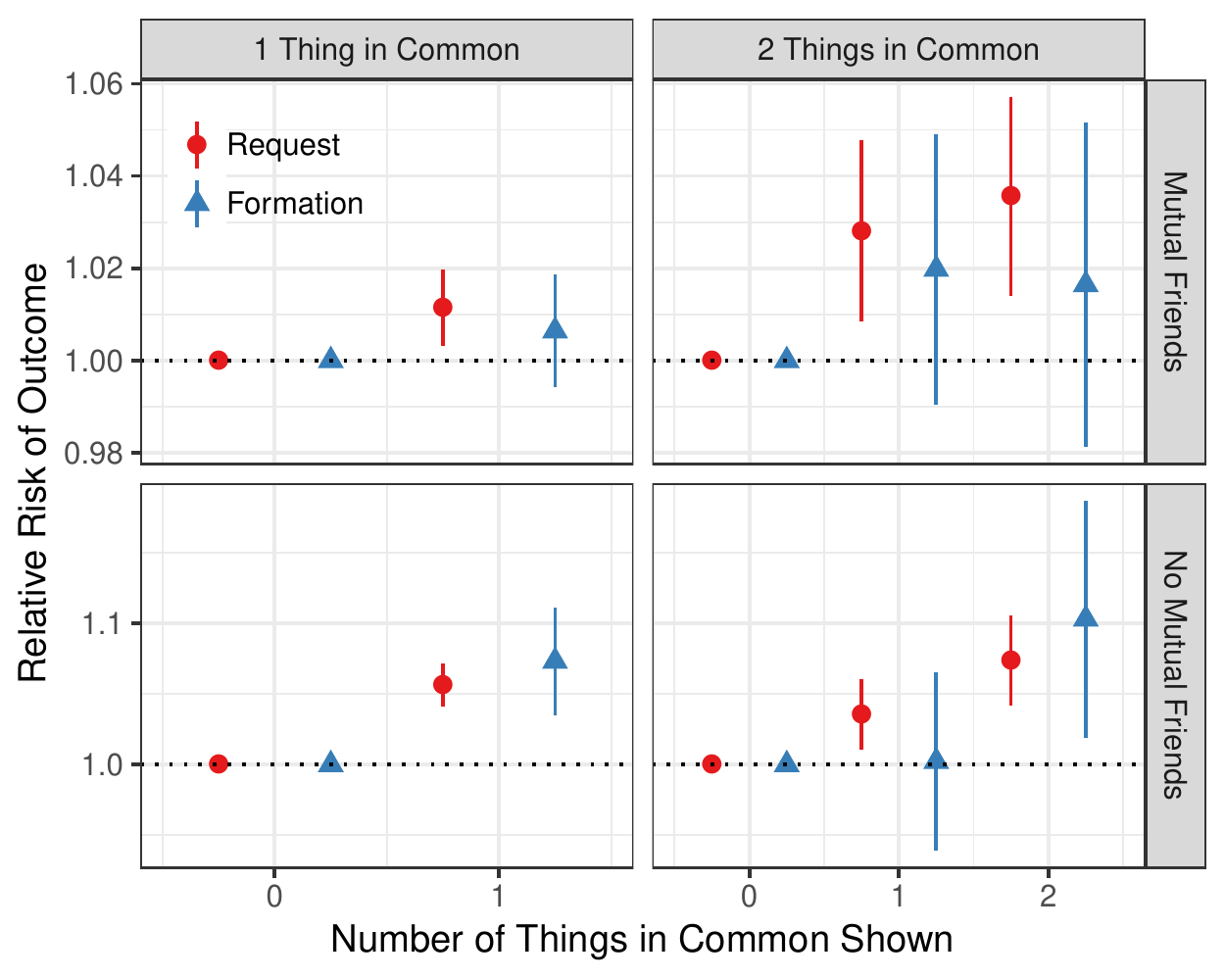}
\caption{\emph{Effects of displaying things in common, for pairs with and without mutual friends.} We group all pairs by the number of things in common as well as the existence of mutual friends. Using pairs with zero things in common shown as a baseline within each group, we find showing more things in common increases the relative risk of friendship formation, but only for people with no mutual friends. 95\% confidence intervals are estimated using a bootstrap clustered on viewer with $500$ bootstrap replicates. \\ \textbf{Note:} for display purpose, the y-axis scale for the upper panel and lower panel is different - the effects for pairs with no mutual friends are much larger}
\label{fig:ntic_grouped}
\end{figure}

Second and more interestingly (Q2), we find displaying things in common can effectively encourage friendship formation between weak ties, i.e. people who share little in common \citep{aral2014tie}. Specifically, We divide the sample into pairs of users with and without mutual friends, as shown in Figure \ref{fig:ntic_grouped}. Displaying TIC can encourage additional friend requests for all viewers, but are especially effective when the viewer and profile does not share any mutual friend (as revealed from the difference in y-scale). The friendship formation outcome follows a similar pattern but are more salient. In contrast to viewer-profile pairs with mutual friends (Figure \ref{fig:ntic_grouped} upper panel), displaying things in common can lead to a large and significant increase in friendship formation rate for pairs without mutual friends (lower panel): 7.3\% relative increase for pairs with 1 TIC, and 10.3\% for pairs with 2 TIC (see detailed estimates in Table \ref{tab:effects2} for relative risk and other estimates). An immediate implication of the above findings is that the proposed information intervention could encourage the formation of weak ties. Displaying things in common may work without structural constraint (i.e. mutual friend) and connect people with even modest similarity (i.e. 1 or 2 things in common), therefore increase diversity of users’ friendship ties and potentially reduce homophily and clustering in the social network.

Third (Q2), we further explore the heterogeneity in the treatment effect, and find that the information intervention works for a wide range of viewers with different demographics and friendship status.  Specifically, we find the effect of displaying things in common is effective for male and female viewers (Figure \ref{fig:hte_viewer_gender}), and also fairly consistent across various gender combinations (Figure \ref{fig:hte_viewer_profile_gender}, e.g. a male browsing a female profile). In addition, we find the effects of displaying things in common is also consistent across viewers with different number of friends (Figure \ref{fig:hte_viewer_fr_ct}\footnote{For results in Figure \ref{fig:hte_viewer_gender}, \ref{fig:hte_viewer_profile_gender} and \ref{fig:hte_viewer_fr_ct}, we also have tables with detailed regression results (available upon request)}), even for the viewers who already have over 500 friends. The above results are aligned with our main findings and suggest that TIC information intervention may be \textit{equally} applied to a large variety of viewers in the social networks and cultivate the formation of weak ties for different subgroups.

\subsection*{\textbf{4.2. Theoretical}}
Besides managerial implication, our findings also provide new insights about the friendship formation process and contributes to the theory on network formation and homophily, in a few ways. First, as discussed earlier, our study provides the first clean experimental evidence on the importance of individuals' preference over thing in common in driving 
friendship formation. We show a direct evidence that after becoming more aware of the shared similarity, the viewer is significantly more likely to send a friend request and form a friendship with the profile. Very interestingly, we also find a variation in the preference over different types of things in common (like, current city, hometown, education, work) and provide more details in Appendix B (Figure~\ref{fig:ntics_1tic}, \ref{fig:ntics_1tic_grouped}).

Second (Q3), we find individuals' preference over TIC plays a more prominent role when there is no mutual friend between the pair. The contrast between treatment effect of displaying TIC for viewer-profiles with and without mutual friends (Figure \ref{fig:ntic_grouped}) suggests a substitution between preference factor and structural factor in friendship formation process, i.e. when a viewer already has mutual friends with the profile, s/he may positively update her/his belief about having TIC with profile. Thus, showing things in common might not reveal much additional information to the viewer thus cannot increase the probability of friendship formation. In other words, mutual friends and things in common may communicate similar or redundant information and thus the signals are not additive. The result is aligned with the belief update process where the viewer's with a prior probability on sharing TIC with the profile is contingent on the existence of mutual friend.

We can better understand the theoretical contributions and mechanisms using a formal model. Specifically, we can characterize $i$’s friend request decision for $j$ with the following equation\footnote{Without loss of generalizability, we focus on the effect of one type of TIC (`a') in the discussion.},

\begin{center}
${Y}_{ij} =  \beta_a \Pr(X_{ia} = X_{ja}) + \beta_a D_{ija} [1 - \Pr({X}_{ia}={X}_{ja})]  + \gamma {X}_{ij}  + {U}_{ij}$
\end{center}

where $\Pr({X}_{ia}={X}_{ja})$ is the viewer $i$’s prior probability of having a shared trait $a$ with profile $j$ when the TIC is not displayed ($D_{ija} = 0$). The coefficient ${\beta}_{a}$ represents the ``importance'' of the TIC to the viewer and captures the preference for the specific type of TIC is in driving the viewer’s friending decision.

When the SN site displays a TIC ($D_{ija} = 1$), the new information leads to a positive update on viewer $i$’s belief about the shared TIC with profile $j$, i.e. $[1 - \Pr({X}_{ia}={X}_{ja})]$, as the viewer is now 100\% sure about the shared trait $a$. The magnitude of the belief update clearly depends on the prior probability of sharing $a$ in common. If a mutual friend is observed between $i$ and $j$, then the viewer $i$’s ex-ante expectation on the likelihood of having a TIC (a) with profile $j$, i.e. $\Pr({X}_{ia}={X}_{ja})$, will be higher, thus the belief update (or information gain: $1 - \Pr({X}_{ia}={X}_{ja})$) from seeing TIC will be smaller. In other words,  $1 - \Pr({X}_{ia}={X}_{ja} | \mbox{mutual friends} = 0) > 1- \Pr({X}_{ia}={X}_{ja} | \mbox{mutual friends} = 1) $. Therefore, we expect the effect of displaying TIC should be larger and more significant for pairs without mutual friends, and the findings confirm such belief update process.

The substitution effect between preference factor and structural factor also suggests another strength of weak ties \citep{granovetter1977strength} but during friendship formation stage: the marginal value of displaying TIC is larger for weak ties in the tie formation stage, as ex-ante the users have lower expectation on having shared attributes thus may gain more from TIC information. This insight is complementary to recent findings in the social influence literature \citep{aral2014tie,bakshy2012social}\footnote{Social influence spread on existing ties is strongest when the tie shares mutual friends and multiple TICs.} and paints a complete picture of the dynamics across tie formation stage and influence spread stage \citep{phan2018evolution,aral2012identifying}. In this way, our experiment findings shed light on the origin and evolution of homophily observed in social networks\footnote{Though preference over similar others is a underlying driver of homophily among weak ties, the preference, by itself, does not reinforce tie formation among people who already share mutual friends thus cannot lead to a significant level of structural homophily. Preference may connect people with different background, but the rest of network evolution might be driven by structural factor such as meeting opportunities and triadic closure}. 

Third (Q3, Mechanism), we further examine the potential mechanism underlying the effect of showing things in common on a viewer's friending behavior. As discussed above, displaying TICs may update an individual's belief about the shared similarity with the profile \citep{goel2010real}, reduce the search cost and lead to a significant information gain. Aligned with belief update process, we expect showing a TIC would be more effective if the information is less expected by the viewer ($\Pr({X}_{ia}={X}_{ja})$) is small). Consider the case of mutual hometown, a viewer $i$ from a smaller city would encounter fewer profiles who share the same hometown and thus ex-ante have a lower prior probability of sharing the same attribute with another user. In other words, the viewer's expectation on a TIC would be small when the TIC is rarely encountered by her/him. Displaying such TIC in turn would lead to a larger increase in friendship formation.\footnote{Another potential mechanism is attention disruption: showing TIC on a profile card would disrupt the monotonicity of seeing many profile cards in a row. Such additional attention may lead to an increase likelihood of friendship formation. The attention disruption explanation suggests that the effect of TIC is stronger in the early stage of browsing, especially during the transition between control and treatment. However, we perform an exploratory analysis on whether the effect of displaying TIC would vary across profiles with different positions in the browsing sequence but did not find any significant pattern. The evidence might indicate that attention disruption is not playing a major role underlying the process. We thank one reviewer for the suggestion.}

\begin{figure}[h!]
\centering
\includegraphics[width=0.7\textwidth]{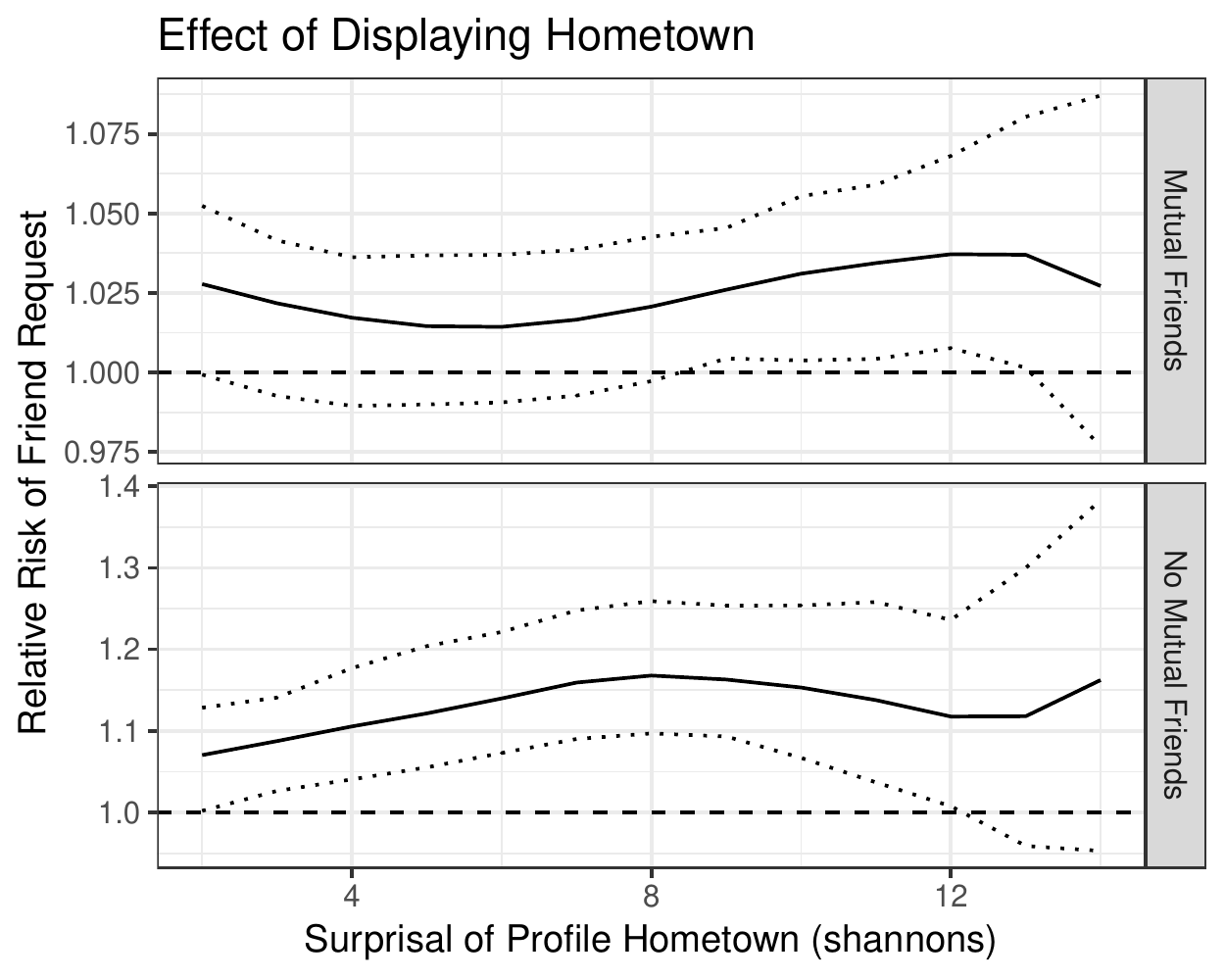}
\caption{\emph{Effect of Displaying Hometown, of Different Surprisal.} We measured the `Surprisal' of the hometown by taking the negative (base 2) log probability of a viewer seeing a profile with that mutual hometown.  Higher bits indicates we would estimate the mutual hometown would be more suprising to the viewer.  To generate smooth curves, we fit LOESS regressions separately for the display and no-display conditions for each of bootstrap replicate.  The dashed lines are the 95\%  confidence intervals for the smoothed estimates of the ratio, computed from a standard errors estimated using a bootstrap clustered on the viewer with $500$ bootstrap replicates.\\\textbf{Note:} for display purpose, the y-axis scale for the upper panel and lower panel is different - the effects for pairs with no mutual friends are much larger}
\label{fig:1tic_ht_surprisal_grouped}
\end{figure}

We follow the information-theoretic framework \citep{shannon1948mathematical} and operationalize the above hypothesis using our data by computing the `Surprisal' of encountering TIC (or `Information gain', $-\log_2(\Pr(X_{ia} = X_{ja}))$, with bits as units for all viewers and attributes in the experiment. One bit is the information gain from the display of a TIC when viewers' prior of observing the TIC is 50\%. Viewers with rare profile attribute, such as uncommon hometown or small current city, encounter profiles who share that TIC information only rarely thus have a lower prior. As a result, the surprisal when the information revealed from displaying the TIC is large.  We compute the relative risk of friend requests as a function of whether the TIC was displayed and the surprisal of that TIC in  Figures~\ref{fig:1tic_ht_surprisal_grouped}
. As shown by the x-scales, sharing a hometown (or current city) with other profiles tends to contain about 2 to 16 bits of information.  For viewer-profile pairs without mutual friends, we observe a positive correlation between the surprisal of the TIC and the effect of displaying that information\footnote{The positive correlation is tapering off at the right end (in the area where surprisal$>$9 shannons). Since the number of observations is much smaller in this area (as revealed from the wider confidence interval), it does not strongly affect the overall trend. The coefficient of surprisal is positive if we fit a linear relationship on the data.}.  For pairs with mutual friends there is no significant effect, indicating that the existence of mutual friends attenuated the effect of TIC information and make it redundant.

In addition, the belief update mechanism and our operationalization based on the information-theory may also guide social network sites to decide on whether and when to display multiple TICs. Showing multiple TICs, as compared to showing one TIC would be valuable only when the additional TIC shown can still bring incremental contribution on the information gain (i.e. `surprisal', $-\log_2(\Pr({X}_{ia}={X}_{ja} | \mbox{other TICs displayed}))$). We find the evidence from the experiment aligned with the argument. As demonstrated in Figure \ref{fig:2tic_cc_ht_surprisal_grouped}, the effect of displaying TIC is stronger when the surprisal brought by displaying both TIC (mutual hometown and current city) is larger than displaying only 1 TIC. SN sites can leverage the principle and our information-theoretic framework (`surprisal') when designing a menu of TIC information for display\footnote{Interestingly, at an aggregate `type' level, the information across different types of TICs are likely to be substitutes for one another. We can identify such relationship by examining the effect of TIC display when the viewer-profile pairs share two TIC (Figures~\ref{fig:2tics_dist}~and~\ref{fig:2tics_effect}). As shown from the 9 panels in Figures~\ref{fig:2tics_effect}, in most scenarios, the effect of displaying two types of things in common is not additive, demonstrating no clear positive complementarity between them is identified. Thus, SN sites could use the information-theoretic framework to guide the optimal display of TIC.}.

\section*{5. Conclusion and Future Research}
Friendship formation is crucial for society in general: it may facilitate information transfer \citep{gee2017paradox}, improve trust \citep{bapna2017repeated}, and broaden diversity \citep{eagle2010network}, yet little is known about how policy makers and SN sites may actively encourage friendship formation, especially for those with relatively little in common. Using a carefully-designed large-scale randomized experiment, our study is among the first to demonstrate the causal effect of displaying things in common on friendship formed. Providing information on shared traits, enabled by big data technology, may significantly reduce the information friction in friendship formation. Such intervention encourages network formation among weak ties and may significant boost friendship diversity. Our experiment provides managerial implications and offer guidelines for SN sites to design effective information intervention in a personalized way. 

The findings also advance our theoretical understanding and contribute to the Marketing, Information Systems and Social Network literature. First, our study provides the first experimental evidence on the role of preference over different TIC in driving network formation.
Moreover, interestingly, we identify and explain a substitution effect between preference factor and structural factor in social network evolution. Finally, our finding is aligned with a rational belief update process: the effect size of showing a certain TIC is in proportion to the `surprise' it creates for the viewer. SN can leverage our information-theoretic framework when designing an optimal information set for display. 

Our study proposes a new experiment design to engineer interpersonal relationship at large-scale and reveals the potential of providing TIC 'information' in facilitating social interactions. We envision future research may extend our study in a few ways. First, our study is conducted in the context of SN sites. Future research may examine the effectiveness of displaying TIC information within organizations (through enterprise social network like Slack, Jive and Yammer, \cite{huang2015structural}), and also in other forms of social interactions, such as hiring, team matching and dating \citep{bapna2016one,gee2017paradox,granovetter1977strength,fisman2006gender}. Second, new types of TIC are emerging on SN sites and researchers could evaluate the effectiveness of those TICs (e.g. location or event check-in). It would be interesting to understand how real-time environment (e.g. at the same event) would encourage friendship formation and interactions. Third, our experiment focuses on the `information' aspect, rather than the relative `importance' in the individuals' preference over different TIC (e.g. hometwon, interest). We believe `information' can reduce friction in social interactions and help individuals better understand each other, while `importance' might be formed based on an individual's experience and largely fixed. 
Future research may leverage exogenous shocks and panel data to understand whether such preference would evolve over time \citep{phan2015natural,benabou2016mindful,bursztyn2017extreme}.
Fourth, we focus on the effect of TIC display on the friendship formation outcome at a viewer-profile level. More research could be done to understand the user behaviors at the viewer level or even network level. For instance, future experiments with randomization at viewer level could help understand whether there exists a ‘saturation’ stage in the users’ need for friendship, i.e. whether users have a quota of friend request or friendship formation that they have in mind. In that case, different friendship may be substitutes to each other in the network formation process. In addition, future research may also leverage simulation methods to understand how network structure and homophily evolves over a longer period of time. 

Finally, our study is among the first to demonstrate the possibility of directly engineering certain preference factors in the process of social network formation. Researchers in Marketing, Information Systems, Economics and Organization Behavior can use our experiment design as a testbed to examine various preference and structural factors underlying interpersonal interactions in network formation. For instance, highlighting certain type of differences (rather than similarities) between two users often would be effective as well, as users may be curious about people who are different from themselves (e.g. sharing a different interest). It is valuable to understand and test what types of differences may be complementary and could be displayed to encourage friendship formation. Our study provides an ideal experiment framework for future inquiry into this broad research agenda on social network formation. 



\bibliographystyle{apalike}
\bibliography{references}

\clearpage
\renewcommand\thesection{\Alph{section}}
\section*{Appendices}
\section{Tables and Figures}
\setcounter{figure}{0} 
\setcounter{table}{0} 
\counterwithin{figure}{section}
\counterwithin{table}{section}

\begin{table}[ht]
\centering
\begin{tabular}{ccc|rrr|cc}
  \hline
\multicolumn{1}{p{1cm}}{Mutual Friends} &
\multicolumn{1}{p{1cm}}{\centering TICs} & 
\multicolumn{1}{p{1.7cm}|}{\centering Displayed TICs} & 
Exposures & Requests & Accepts & 
\multicolumn{1}{p{1.7cm}}{\centering $\Pr(\mbox{Request})$ x 100} & 
\multicolumn{1}{p{1.7cm}}{\centering $\Pr(\mbox{Accept})$ x 100}
\\
  \hline
  no &  1 &  0 & 6.8e+06 & 3.4e+04 & 5.8e+03 & 0.49 & 17.27 \\ 
     &    &  1 & 6.8e+06 & 3.6e+04 & 6.2e+03 & 0.52 & 17.49 \\ 
\hline     
  no &  2 &  0 & 2.2e+06 & 1.0e+04 & 1.4e+03 & 0.45 & 14.22 \\ 
     &    &  1 & 4.5e+06 & 2.1e+04 & 2.9e+03 & 0.47 & 13.83 \\ 
     &    &  2 & 2.2e+06 & 1.1e+04 & 1.6e+03 & 0.48 & 14.75 \\ 
\hline     
  no &  3 &  0 & 6.8e+04 & 8.4e+02 & 1.1e+02 & 1.24 & 12.54 \\ 
     &    &  1 & 2.1e+05 & 2.7e+03 & 3.4e+02 & 1.33 & 12.63 \\ 
     &    &  2 & 2.1e+05 & 2.9e+03 & 3.4e+02 & 1.43 & 11.66 \\ 
     &    &  3 & 6.9e+04 & 9.4e+02 & 9.8e+01 & 1.37 & 10.44 \\ 
\hline     
\hline
  no &    &    & 2.3e+07 & 1.2e+05 & 1.9e+04 & 0.51 & 15.90 \\
\hline
\hline
  yes &  1 &  0 & 1.1e+07 & 1.2e+05 & 5.2e+04 & 1.14 & 41.86 \\ 
      &    &  1 & 1.1e+07 & 1.3e+05 & 5.2e+04 & 1.15 & 41.68 \\ 
\hline      
  yes &  2 &  0 & 8.7e+05 & 1.7e+04 & 6.9e+03 & 1.91 & 41.60 \\ 
      &    &  1 & 1.7e+06 & 3.4e+04 & 1.4e+04 & 1.96 & 41.27 \\ 
      &    &  2 & 8.7e+05 & 1.7e+04 & 7.0e+03 & 1.97 & 40.74 \\ 
\hline      
  yes &  3 &  0 & 1.6e+05 & 2.6e+03 & 7.3e+02 & 1.65 & 28.05 \\ 
      &    &  1 & 4.8e+05 & 8.0e+03 & 2.3e+03 & 1.68 & 28.58 \\ 
      &    &  2 & 4.8e+05 & 8.1e+03 & 2.3e+03 & 1.70 & 28.98 \\ 
      &    &  3 & 1.6e+05 & 2.8e+03 & 8.1e+02 & 1.79 & 28.51 \\ 
   \hline
   \hline
  yes &     &     & 2.7e+07 & 3.4e+05 & 1.4e+05 & 1.28 & 40.82 \\
  \hline
  \hline
Total &     &     & 5.0e+07 & 4.6e+05 & 1.6e+05 & 0.92 & 34.35 \\
  \hline
\end{tabular}
\caption{\emph{Summary Statistics} \newline Overall the experiment includes about 50 million viewer-profile pairs.  A large fraction of the sample comes from viewer-profile pairs with one TIC (~73\%) and two TIC (~25\%). Among all pairs, about 23 million share no mutual friends (upper panel) and 27 million pairs share mutual friends (lower panel).  The friend request rate is correlated with the number of TICs, while the accept rate is negatively correlated with it.}
\label{tab:summary_stats}
\end{table}

\begin{table}[ht]
\centering
\begin{tabular}{cc|rrcc|rrcc}
  \hline
TICs & 
\multicolumn{1}{p{1.7cm}|}{\centering Displayed TICs} &
\multicolumn{4}{c|}{Friend Requests} &
\multicolumn{4}{c}{Friend Formations} \\
\multicolumn{2}{p{2.5cm}|}{} &
RR & Effect & s.e. & p-value & 
RR & Effect & s.e. & p-value \\ 
\hline
1 & 1 & 1.021 & 1.89 & (0.34) & 0.00 & 1.013 & 0.43 & (0.20) & 0.01 \\ 
\hline
2 & 1 & 1.032 & 2.70 & (0.65) & 0.00 & 1.018 & 0.48 & (0.36) & 0.10 \\ 
  & 2 & 1.052 & 4.41 & (0.79) & 0.00 & 1.034 & 0.91 & (0.45) & 0.03 \\ 
\hline  
3 & 1 & 1.034 & 5.21 & (2.88) & 0.03 & 1.050 & 1.83 & (1.36) & 0.09 \\ 
  & 2 & 1.061 & 9.31 & (2.85) & 0.00 & 1.071 & 2.62 & (1.32) & 0.03 \\ 
  & 3 & 1.089 & 13.50 & (3.70) & 0.00 & 1.081 & 2.98 & (1.84) & 0.04 \\ 
   \hline
\end{tabular}
\caption{\emph{Effect of Displaying TICs on Friend Requests and Formation} \newline RR stands for relative risk and is calculated using the relative ratio in means compared to the baseline group (i.e. viewer-profile pairs with the same number of actual TICs but zero TICs at display). RR corresponds to the y-scale in Figure \ref{fig:ntics}. Effect is measured using the absolute difference in means compared to showing zero TICs (baseline) and denominated here in basis points for readability.  Standard errors are estimated using a clustered bootstrap on viewer with $500$ bootstrap replicates.}
\label{tab:effects1}
\end{table}

\begin{table}[ht]
\centering
\begin{tabular}{ccc|ccc|ccc}
  \hline
\multicolumn{1}{p{1.5cm}}{\centering Mutual Friends} & 
TICs & 
\multicolumn{1}{p{1.7cm}|}{\centering Displayed TICs} &
\multicolumn{3}{c|}{Friend Requests} &
\multicolumn{3}{c}{Friend Formations} \\
\multicolumn{3}{p{2.5cm}|}{} &
RR & s.e. & p-value & 
RR & s.e. & p-value \\ 
  \hline
  no & 1 & 1 & 1.056 & (0.38) & 0.00 & 1.073 & (0.16) & 0.00 \\ 
\hline  
  no & 2 & 1 & 1.036 & (0.56) & 0.00 & 1.002 & (0.21) & 0.49 \\ 
     &   & 2 & 1.073 & (0.71) & 0.00 & 1.103 & (0.26) & 0.00 \\ 
\hline     
  no & 3 & 1 & 1.075 & (5.37) & 0.04 & 1.119 & (1.66) & 0.16 \\ 
     &   & 2 & 1.145 & (5.4) & 0.00 & 1.126 & (1.68) & 0.15 \\ 
     &   & 3 & 1.099 & (5.95) & 0.02 & 0.961 & (2.10) & 0.64 \\ 
\hline  
  yes & 1 & 1 & 1.011 & (0.49) & 0.00 & 1.006 & (0.30) & 0.16 \\ 
\hline  
  yes & 2 & 1 & 1.028 & (1.87) & 0.00 & 1.020 & (1.18) & 0.08 \\ 
      &   & 2 & 1.036 & (2.05) & 0.00 & 1.016 & (1.42) & 0.19 \\ 
\hline      
  yes & 3 & 1 & 1.022 & (3.61) & 0.17 & 1.041 & (1.84) & 0.16 \\ 
      &   & 2 & 1.035 & (3.67) & 0.06 & 1.066 & (1.76) & 0.04 \\ 
      &   & 3 & 1.086 & (4.66) & 0.00 & 1.100 & (2.42) & 0.02 \\ 
\hline
\end{tabular}
\caption{\emph{Effect of Displaying TICs on Friend Requests and Formation (Decomposed by Whether View-Profile Pairs Share Mutual Friends)} \newline RR stands for relative ratio and is calculated using the relative lift in means compared to the baseline group (i.e. viewer-profile pairs with the same number of actual TICs but zero TICs at display). Standard errors are estimated using a clustered bootstrap on viewer with $500$ bootstrap replicates.}
\label{tab:effects2}
\end{table}

\begin{figure}[h!]
\centering
\includegraphics[width=0.4\textwidth]{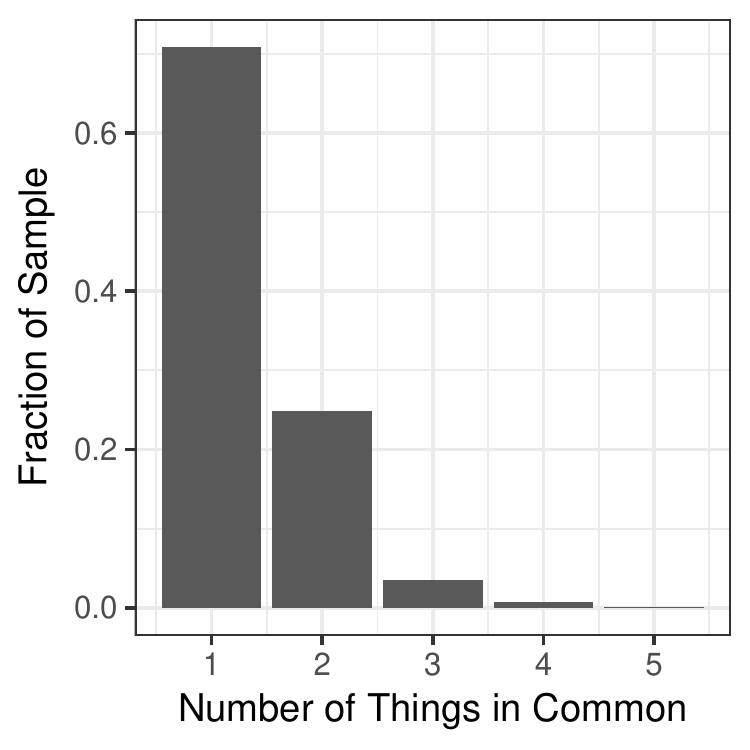}
\caption{\emph{Distribution.} We group all pairs by the number of actual things in common shared between the viewer-profile and draw its distribution. Most viewer-profile pairs have one thing in common.}
\label{fig:ntics_dist}
\end{figure}

\begin{figure}[h!]
\centering
\includegraphics[width=0.7\textwidth]{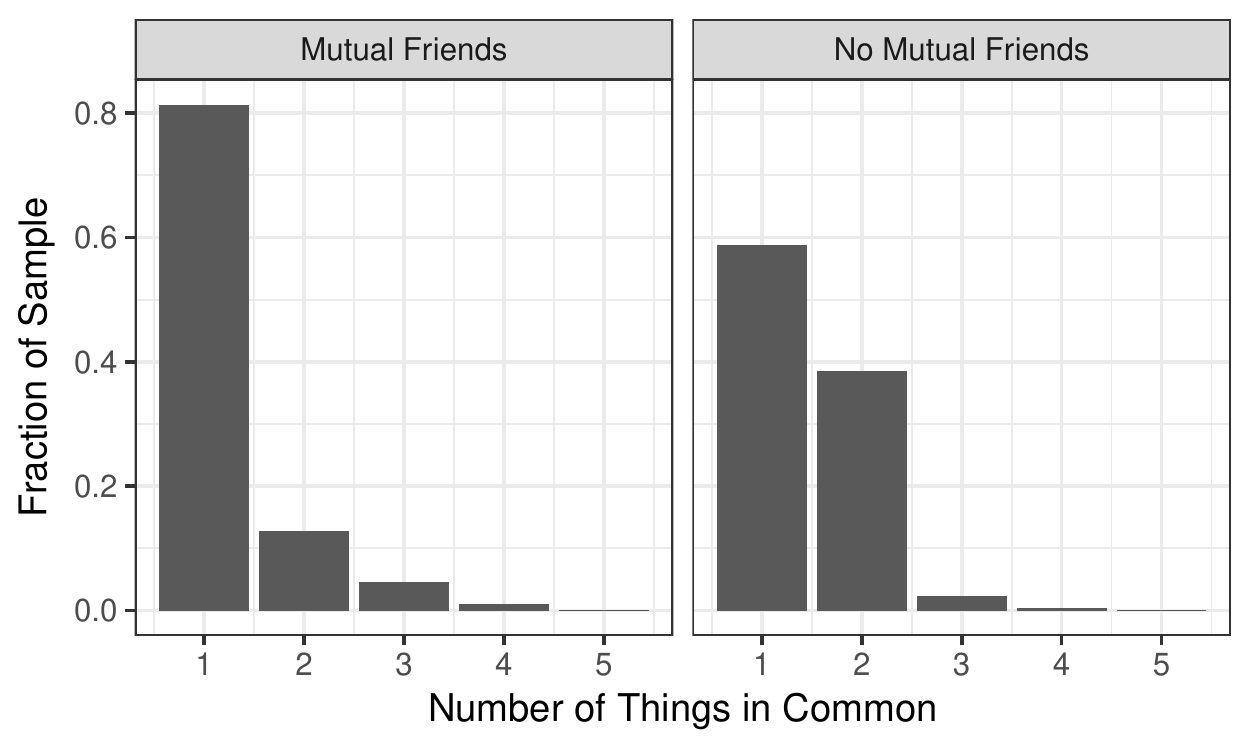}
\caption{\emph{Distribution.} We group all pairs by the number of actual things in common shared between the pairs and draw separate distribution for pairs with and without mutual friend.}
\label{fig:ntics_dist_grouped}
\end{figure}

\begin{figure}[h]
\centering
  \begin{subfigure}[b]{0.40\textwidth}
    \includegraphics[width=\textwidth]{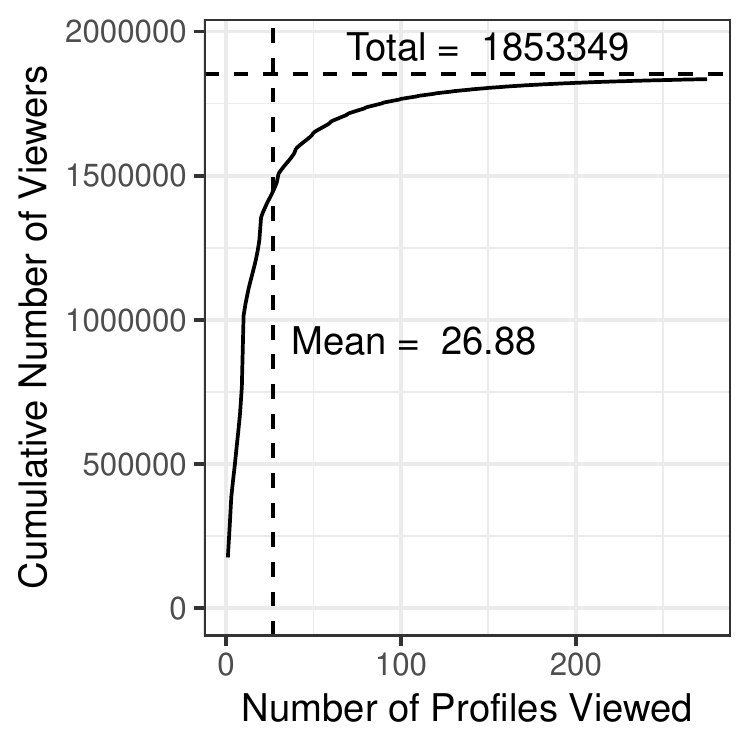}
    \label{fig:dist_profile_per_viewer}
  \end{subfigure}
  \begin{subfigure}[b]{0.40\textwidth}
    \includegraphics[width=\textwidth]{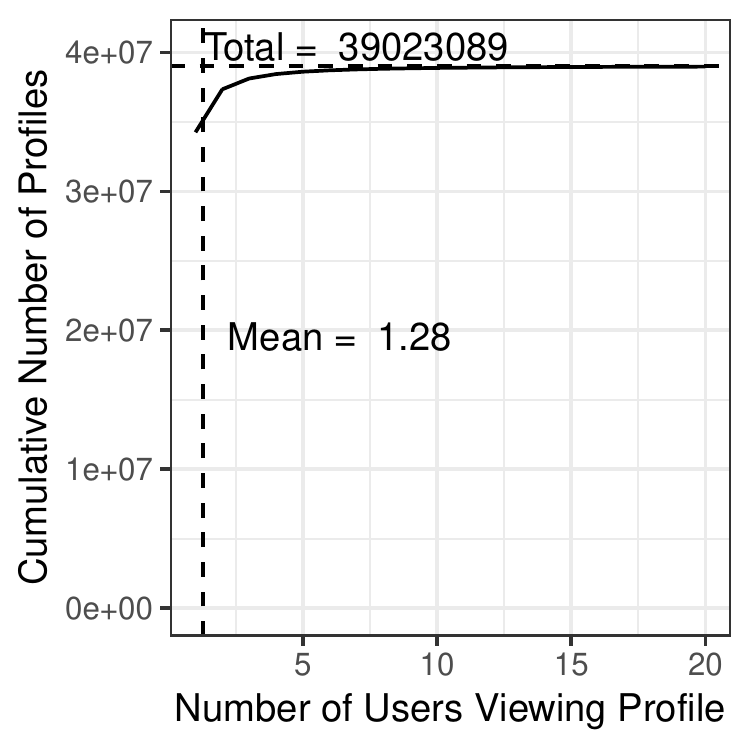}
    \label{fig:dist_viewer_per_profile}
  \end{subfigure}
\caption{\emph{Distribution of profiles per viewer (left) and Distribution of viewers per profile (right).}}
\label{fig:dist_profile_viewer}
\end{figure}

\begin{figure}[h!]
\centering
\includegraphics[width=0.8\textwidth]{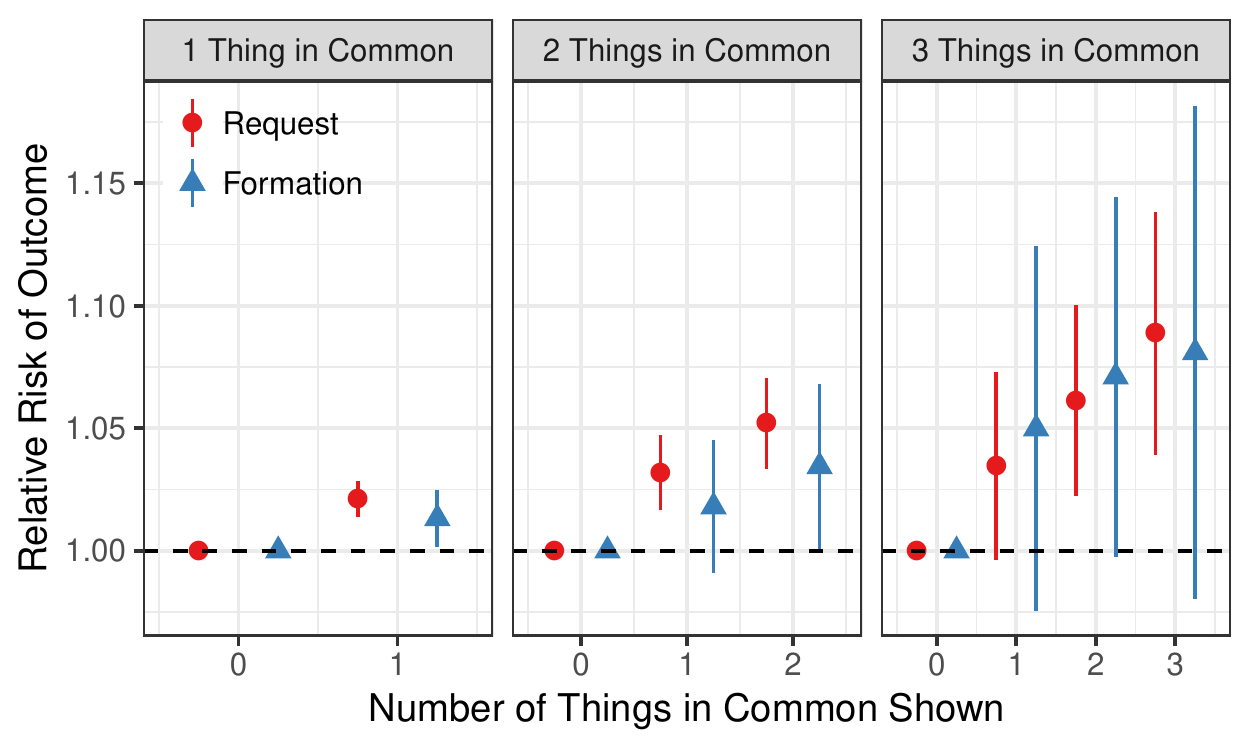}
\caption{\emph{Effects of displaying things in common.} Same as Figure \ref{fig:ntics} with an additional panel for pairs with 3 TIC}
\label{fig:ntics_3}
\end{figure}

\begin{figure}[h!]
\centering
\includegraphics[width=0.8\textwidth]{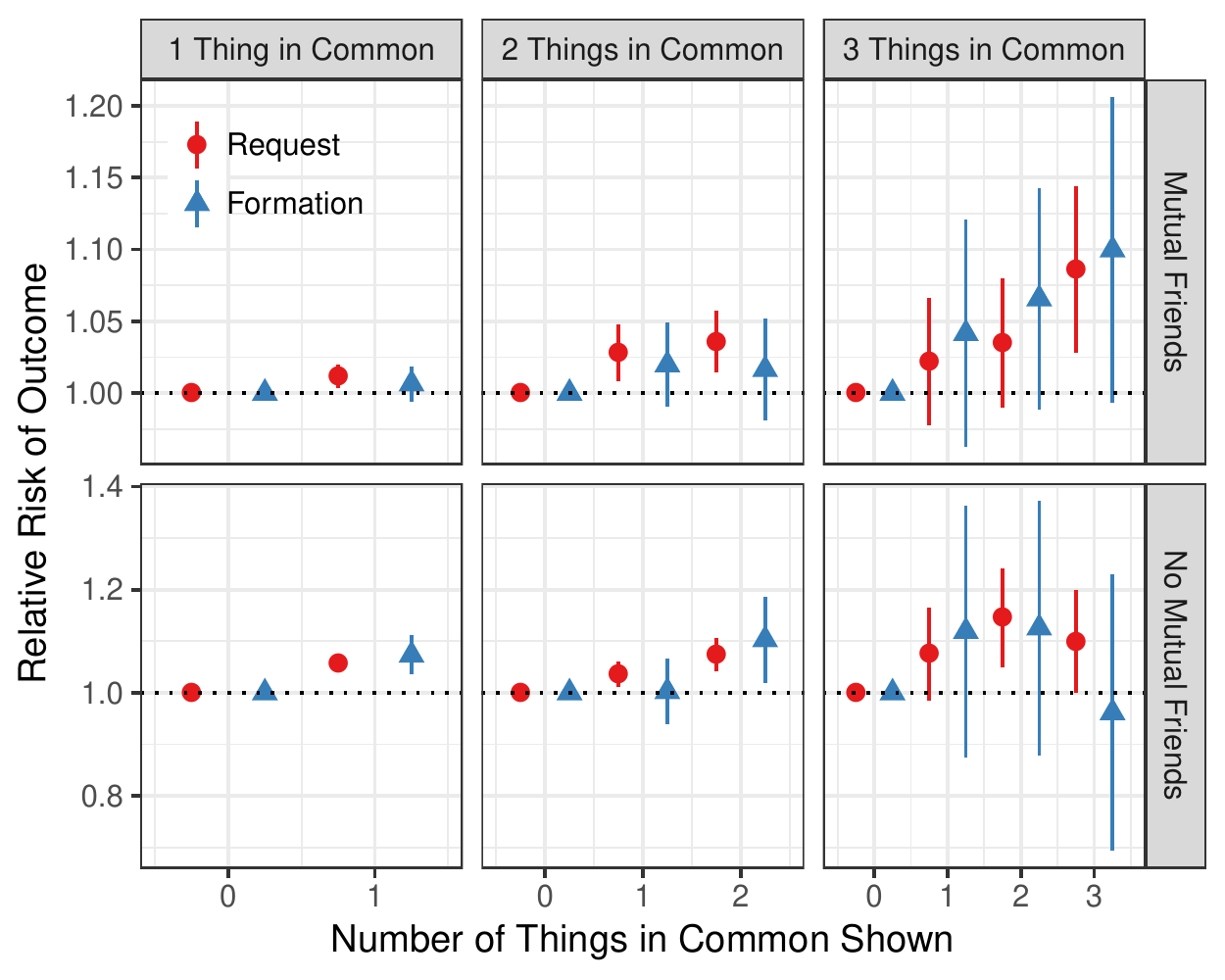}
\caption{\emph{Effects of displaying things in common, for pairs with and without mutual friends.} Same as Figure \ref{fig:ntic_grouped} with an additional panel for pairs with 3 TIC}
\label{fig:ntic_grouped_3}
\end{figure}

\begin{figure}[h!]
\centering
\includegraphics[width=0.6\textwidth]{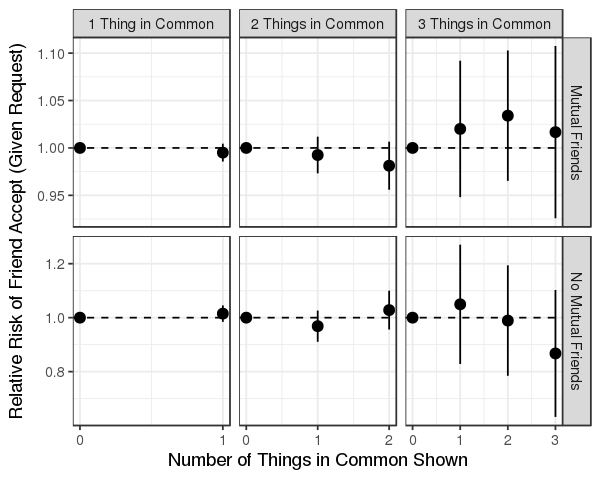}
\caption{\emph{Acceptance Rate Conditional on Request}}
\label{fig:conditional_acceptance_rate}
\end{figure}

\begin{figure}[h!]
\centering
\includegraphics[width=0.7\textwidth]{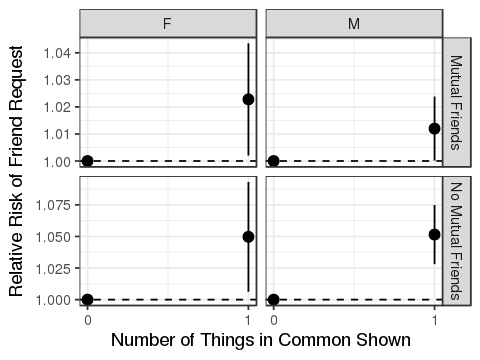}
\caption{\emph{Effect of Displaying Thing in common on Friend Request, for male viewers and female viewers.}}
\label{fig:hte_viewer_gender}
\end{figure}

\begin{figure}[h]
\centering
\includegraphics[width=0.8\textwidth]{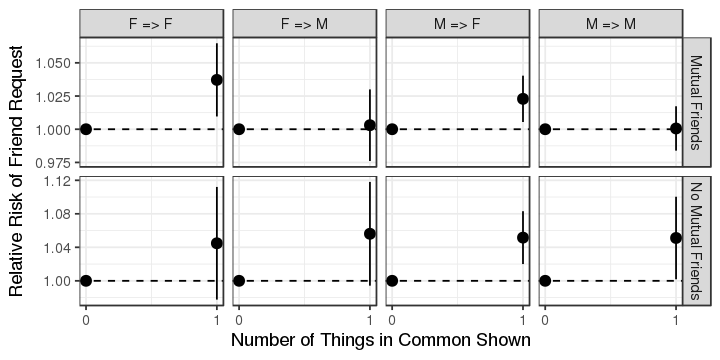}
\caption{\emph{Effect of Displaying Thing in common on Friend Request, for gender combinations of viewers and profiles.}}
\label{fig:hte_viewer_profile_gender}
\end{figure}

\begin{figure}[h]
\centering
\includegraphics[width=0.8\textwidth]{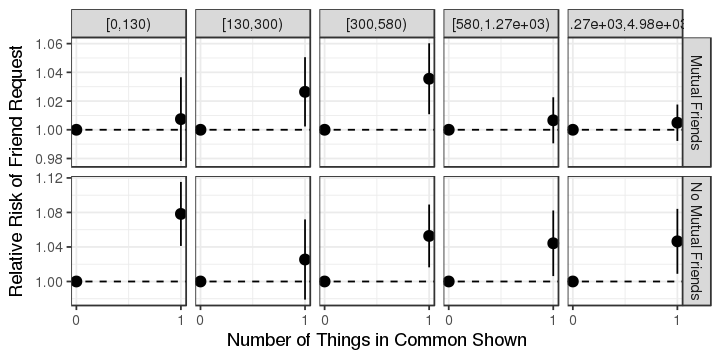}
\caption{\emph{Effect of Displaying Thing in common on Friend Request, for viewers with different number of friends.}}
\label{fig:hte_viewer_fr_ct}
\end{figure}


\begin{figure}[h!]
\centering
\includegraphics[width=0.7\textwidth]{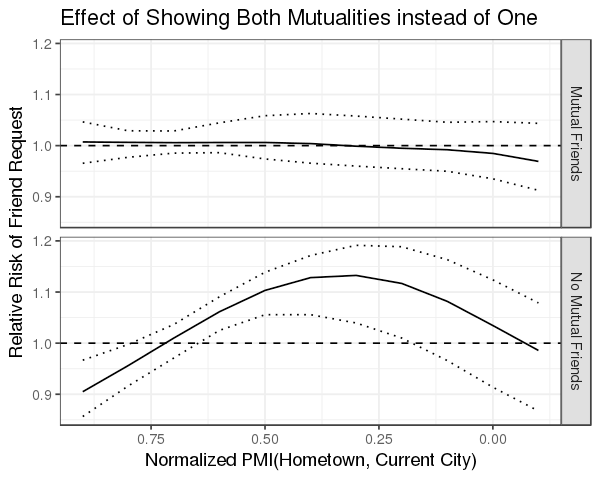}
\caption{\emph{Incremental Effect of displaying both things in Common, as compared to displaying one.} Following information theory, we measured the incremental information (`surprisal') of displaying two TIC (mutual hometown and current city), as compared displaying one, using normalized pointwise mutual information (PMI). We calculate normalized PMI using its formal definition \citep{church1990word}: the surprisal ($-\log_2(\Pr(X_{ia} = X_{ja}))$) of displaying mutual hometown plus the surprisal of displaying current city minus the joint surprisal of displaying both mutual current city and hometown, normalized by joint surprisal. The PMI measure, ranging from +1 to -1, captures the co-occurrence likelihood and is symmetric with respect to the two TICs. An PMI with +1 indicates 100\% likelihood of co-occurrence, thus displaying both mutual hometown and current city would reveal little additional information or `surprisal'. In contrast, a low PMI indicates a small likelihood of co-occurrence thus displaying both TIC would would be more surprising to the viewer given that they already know one of the TIC). The empirical finding is aligned with such mechanism and further confirms the belief update process underlying the effectiveness of displaying TIC. To generate smooth curves, we fit LOESS regressions separately for the display 2 TIC and display 1 TIC conditions for each of 500 bootstrap replicates clustered on viewer.  The dashed lines are the 95\% bootstrap confidence intervals for the smoothed estimates of the ratio.}
\label{fig:2tic_cc_ht_surprisal_grouped}
\end{figure}

\begin{figure}[h]
\centering
\includegraphics[width=0.8\textwidth]{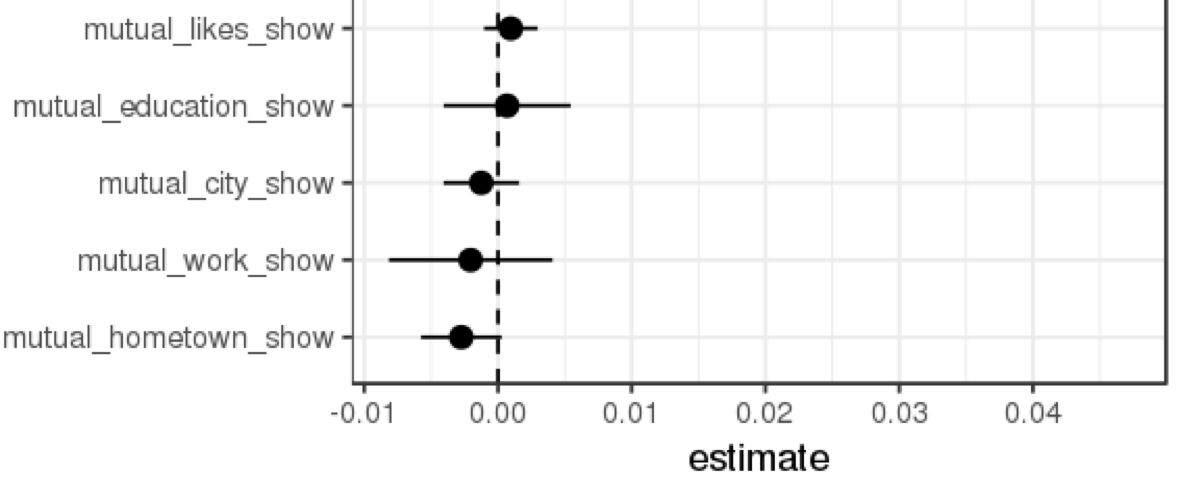}
\caption{\emph{Comparison of Post-friendship Interaction score between friendship ties in control versus treatment group.} We categorize all the friendship ties formed in our experiment into two groups based on whether certain TIC is shown during friendship formation process, and then calculate the difference between the groups in the volume of their post-friendship interactions (like, comment, messaging). We do not find any significant difference in the post-friendship interactions between the control and treatment group.}
\label{fig:post_interact}
\end{figure}

\begin{figure}[h]
\centering
\includegraphics[width=0.8\textwidth]{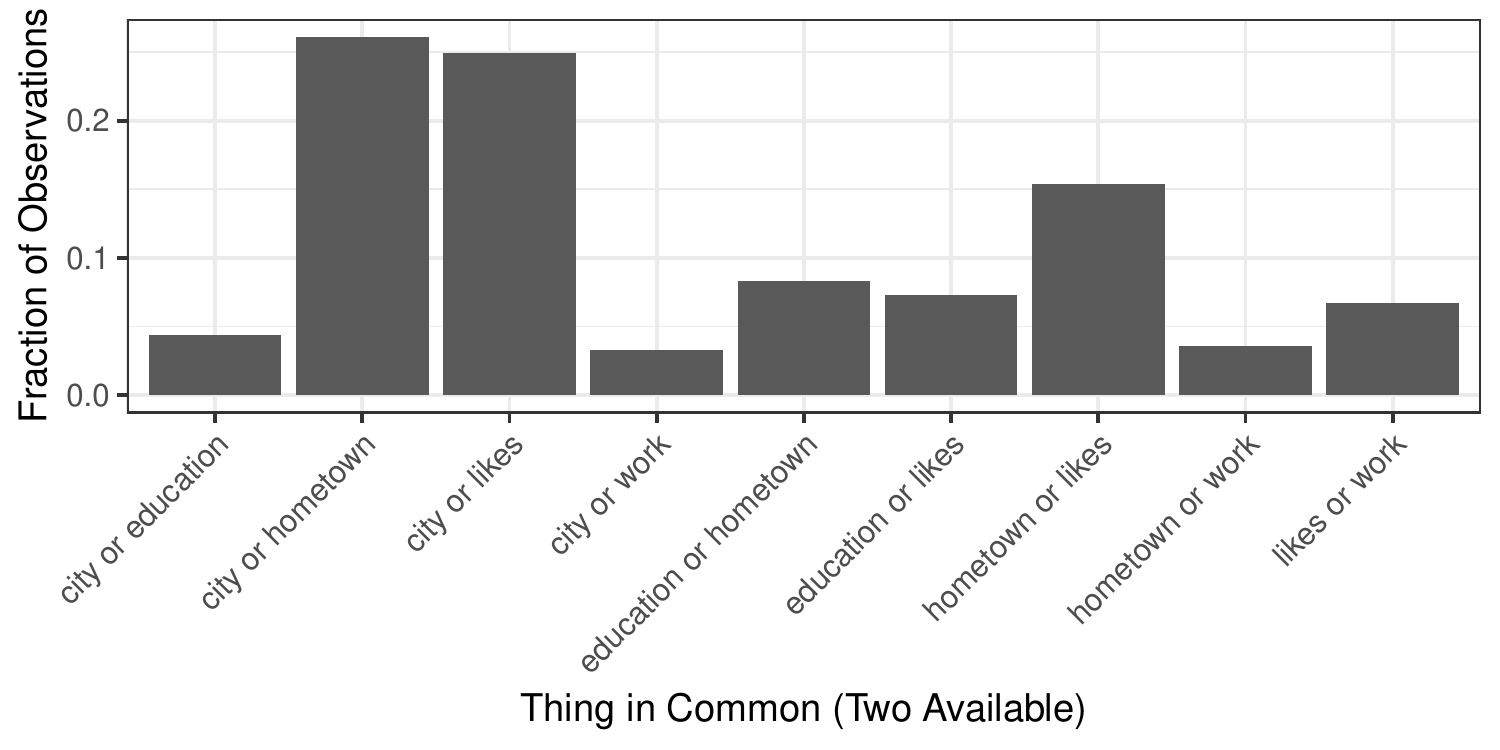}
\caption{\emph{Distribution of things in common when the viewer-profile pair share two things in common.} Among the large number of combinations between the five types of TIC (mutual city, like, hometown, education, work), we only show distributions for the nine most commonly seen combinations. The remaining combinations consist of a very small portion of our sample.}
\label{fig:2tics_dist}
\end{figure}

\begin{figure}[h]
\centering
\includegraphics[width=0.8\textwidth]{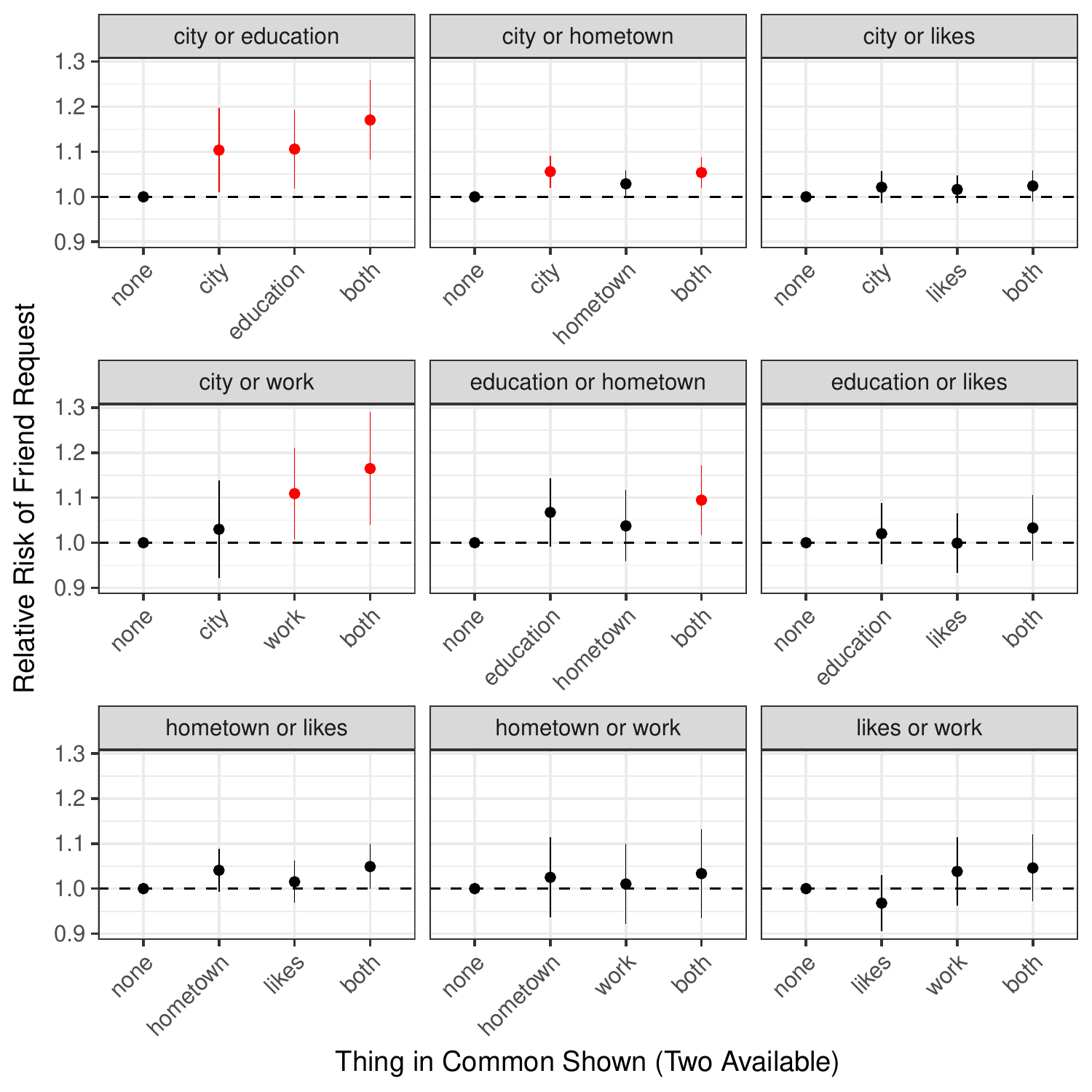}
\caption{\emph{Effects of showing zero, one, or both things in common when the viewer-profile pair share two things in common.} Among the large number of combinations between the five types of TIC (mutual city, like, hometown, education, work), we only show the effect of TIC display for the nine most commonly seen combinations. The remaining combinations consist of a very small portion of our sample.}
\label{fig:2tics_effect}
\end{figure}

\pagebreak
\section{Effect of Displaying Different Types of Things in Common}
\setcounter{figure}{0} 
\counterwithin{figure}{section}

The availability of multiple types of things in common in our experiment also enables us, for the first time, to investigate the relative importance of a range of TIC documented in the literature \citep{mcpherson2001birds} in the same context. Previous studies on homophily have separately documented different things in common in social networks \citep{ibarra1992homophily, currarini2010identifying}, but never compared the causal effect of them within the same context.

\begin{figure}[h]
\centering
  \begin{subfigure}[b]{0.4\textwidth}
    \includegraphics[width=\textwidth]{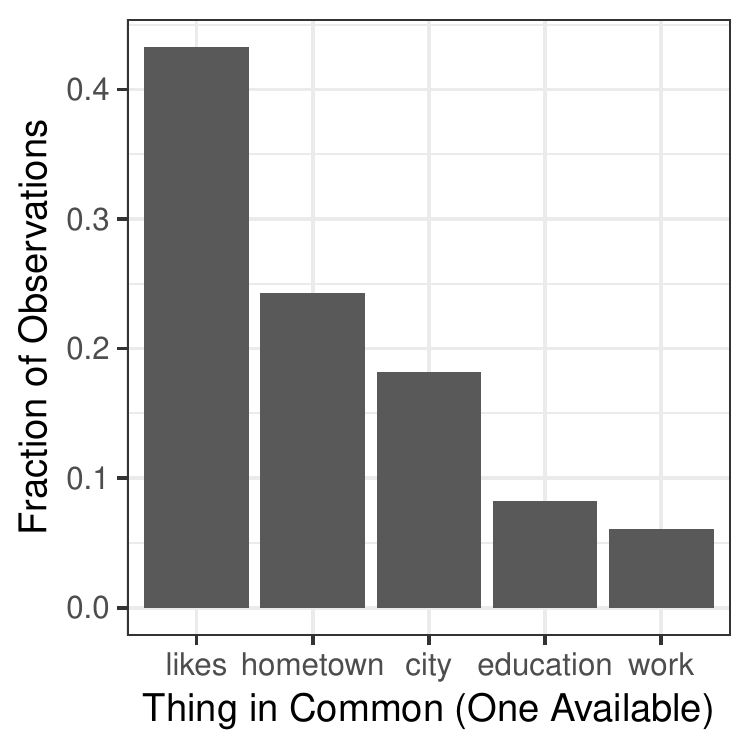}
    \label{fig:1tic_dist}
  \end{subfigure}
  \begin{subfigure}[b]{0.4\textwidth}
    \includegraphics[width=\textwidth]{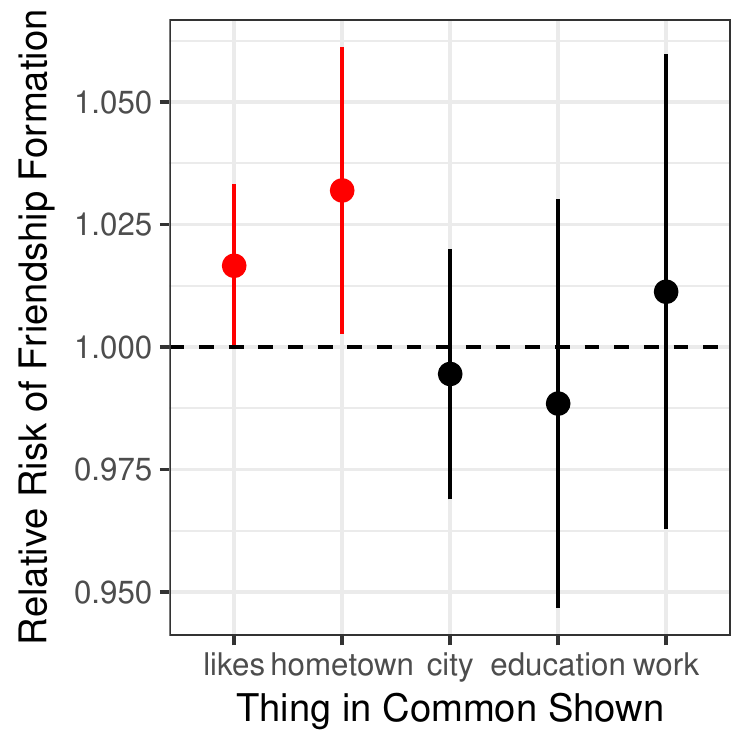}
    \label{fig:1tic_effect}
  \end{subfigure}
\caption{\emph{Distribution and effects of showing specific type of things in common when the viewer-profile pair share one thing in common.}  Using zero things in common as a baseline, showing shared likes and hometown significantly increases the relative risk of friendship formation, while we cannot reject the null hypothesis that showing a shared city or education causes any change. 95\% confidence intervals are computed from standard errors estimated using a bootstrap clustered on viewer with $500$ bootstrap replicates.}
\label{fig:ntics_1tic}
\end{figure}

We find a variation in the effect of showing different types of things in common (Figure~\ref{fig:ntics_1tic}). We focus on presenting the results for viewer-profile pairs with only one thing in common, as they represent the majority of the sample (Figure~\ref{fig:ntics}). There are three categories of things in common: past experience (mutual hometown, mutual education, mutual work), current context (mutual city), common interests (overlap of page like). As shown in Figure~\ref{fig:ntics_1tic}, certain type of things in common, such as mutual likes, city and hometown, may lead to a larger increase in friendship formation than other types of things in common (mutual education). Previous studies on homophily has separately documented different type of things in common across different situations \citep{ibarra1992homophily,currarini2010identifying,thelwall2009homophily}, but never compared the relative importance of them within the same context. Similar to previous studies on social influence \citep{aral2014tie}, we find that social factors such as current context (e.g. mutual city) and past experience (mutual hometown) are associated with a large impact on friendship formation. However, we also find that showing overlap of interest may also be effective: a finding that has not been established before. In this way, we contribute to the literature on the role of different type of things in common (or similarity) in facilitating interpersonal interaction \citep{berscheid1998attraction}.

\clearpage

\begin{figure}[h!]
\centering
\includegraphics[width=0.7\textwidth]{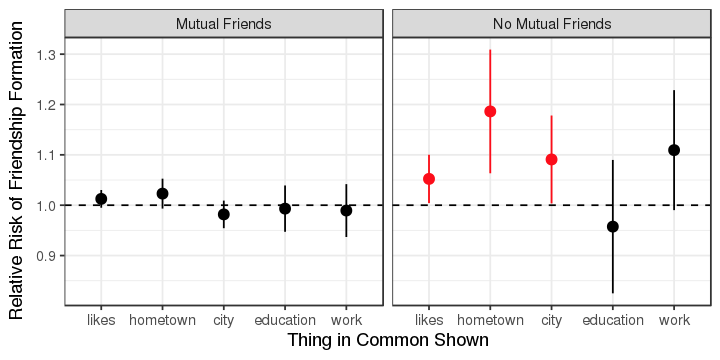}
\caption{\emph{Effect of Specific Type of Thing in common, for pairs with and without mutual friends.} We group all pairs with one thing in common by the specific things in common as well as the existence of mutual friends. Using pairs with zero things in common shown as a baseline within each group, we find showing certain thing in common increases the relative risk of friendship formation, but only for people with no mutual friends and for shared likes, hometown and city. 
}
\label{fig:ntics_1tic_grouped}
\end{figure}

\pagebreak
\section{Methodological: Correlation from Observational Data is Opposite to the Causal Effect of TIC}
\setcounter{figure}{0} 
\counterwithin{figure}{section}

\begin{figure}[h!]
\centering
\includegraphics[width=0.7\textwidth]{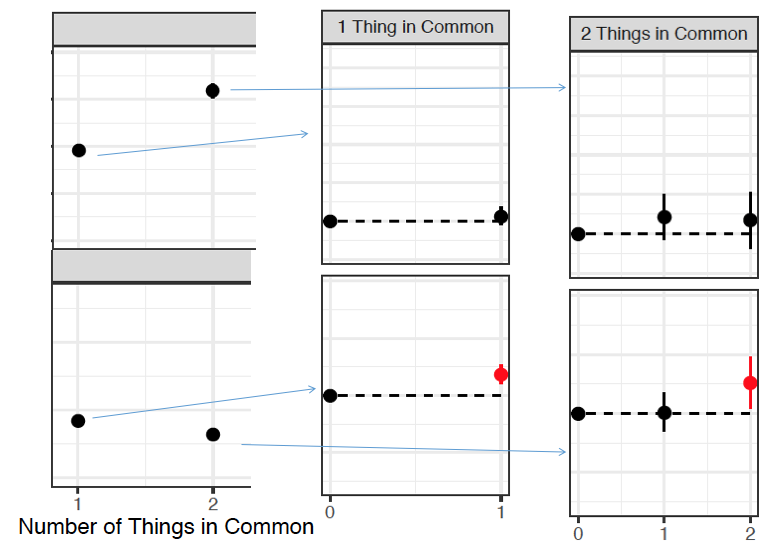}
\caption{\emph{Correlation vs Causal Effect of TIC on friendship formation.} Correlation between number of things in common in a pair and the friendship formation rate (Left) is biased and opposite to the true causal effect of showing things in commons (middle and right).}
\label{fig:corr_vs_causal}
\end{figure}

Finally, we want to demonstrate the importance and necessity of using a randomized experiment approach to identify causal effect of showing TIC. We explicitly compare the estimates from our experiment with those from a direct correlation analysis using observational data to highlight the advantage of our method. In Figure \ref{fig:corr_vs_causal}, we can see that interestingly the correlation between number of things in common and friendship formation is in the opposite direction from the identified causal effect of showing things in common: for viewer-profile pairs with mutual friends, the observed correlation suggests a very positive and significant effect of showing more things in common, i.e. pairs with more TIC is more likely to form friendship, whereas the true causal effect of TIC based on experiment results is zero;  for viewer-profile pairs with no mutual friends, the observed correlation suggests a zero or even slightly negative relationship between more TIC and friendship formation, whereas the true causal effect of showing TIC is large and significant. This contrast further highlights the importance of using a randomized experiment for causal inference. As shown above, using the observed correlation in the secondary data may significantly bias the effect of displayed things in common. We can cleanly identify the causal effect from correlation only by using a proper randomization.

\clearpage
\pagebreak
\section{Theoretical Contribution: Identifying the Role of Individual Preference in Things in Common (versus Structural Factor) in Network Formation}
\setcounter{figure}{0} 
\counterwithin{figure}{section}

Besides practical importance, our findings from a carefully designed large-scale randomized experiment may contribute to social network literature and improve our \textbf{theoretical} understanding of the origin of network formation and homophily, in a few ways.

First, while strategic network formation have been studied in detail using analytical models \citep{bala2000noncooperative,jackson2008social} and simulations \citep{phan2018evolution}, empirical work designed to test theories and examine the drivers of network formation is still in their infancy. Most of the empirical works on network formation \citep{currarini2010identifying,mele2017structural} impose strong functional assumptions and use network structure for identification\footnote{One exception is \cite{phan2015natural}, in which the authors carefully design a long-term natural experiment of friendship formation and social dynamics in the aftermath of a natural disaster.}. To the best of our knowledge, our study is the first randomized field experiment to directly engineer the drivers of network formation (i.e. viewer's information about the alter as in our case). Our experiment design and framework provides a new way to examine various factors underlying network formation and the relative importance between them, in real networks and at a very large scale. 

Second, 
as documented by a broad history of literature on homophily in social network \citep{mcpherson2001birds}, an individual is more likely to form a friendship tie with someone that is similar to her/him in demographic and behavioral attributes \citep{centola2015social,ameri2017structural, goel2013predicting}. However, the origin of homophily is far from clear \citep{kossinets2009origins,currarini2010identifying}. Recent literature hypothesizes that the homophily pattern in social network might be partially driven by individuals' preference over similar others (termed as preference or choice homophily \cite{currarini2009economic}), as opposed to the structural factors. However, as far as we know, no study has provided clear empirical evidence on whether and when preference for TIC would drive network formation. As the literature repeatedly acknowledge \citep{currarini2010identifying,phan2015natural}, such lack of insights is because preference factor and structural factor in network formation process are inherently confounded in observational data and extremely hard to disentangle: pairs with similar attributes are always more likely to have mutual friends. We address the challenge by designing a experiment to exogenously vary the prominence of things in common displayed during friendship formation in a real social network. The randomization is independent of structural factor therefore allows us, for the first time, to demonstrate the importance of preference in driving friendship formation. Third, we examine the relationship between preference and structural factor in network formation and find that
preference over TIC matters in the absence of structural factor (mutual friend) thus can be leveraged to encourage formation of weak ties.

Finally and importantly, the variation in the treatment effect within each type of TIC (e.g. sharing mutual hometowns of different size) sheds light on the underlying mechanisms: aligned with belief update process \citep{ely2015suspense}, displaying TIC is more effective when the information shown is less expected, and the effect size of a certain TIC is in proportion to the 'surprise' it creates to the viewer (measured by bits in information theory). Our study is among the first to understand network formation and homophily from a belief update perspective. SN sites can use our information-theoretic principles to optimally select specific things in common to highlight in online friendship formation process. The information-theoretic framework may also be extended to guide the personalized design of user profile for each viewer-profile interaction.

\end{document}